\documentclass[12pt]{iopart}
\expandafter\let\csname equation*\endcsname\relax
\expandafter\let\csname endequation*\endcsname\relax
\usepackage{amsfonts}
\usepackage{amsmath}
\usepackage{amssymb}
\usepackage{mathabx}
\usepackage{bm}
\usepackage{graphicx}
\usepackage[utf8]{inputenc}
\usepackage[T1]{fontenc}

\usepackage{siunitx}
\usepackage{graphicx}
\usepackage{color}
\usepackage{tikz}

\newcommand{\withlabel}[3][.2, -.5]{%
\begin{tikzpicture}
\node[inner sep=0pt] (pct) at (0,0)
    {#3};
\draw (pct.north west)+(#1) node {\footnotesize\textbf{#2}};
\end{tikzpicture}}

\def\eqref#1{(\ref{#1})}

\def\gbar{\bar{g}}
\def\Hbar{\bar{\mathrm{H}}}

\begin{document}

\title{Analysis of the timing of freely falling antihydrogen}
\author{Olivier Rousselle$^1$, Pierre Clad\'e$^1$, Sa\"ida Guellati-Kh\'elifa$^{1-2}$, Romain Gu\'erout$^1$ and Serge Reynaud$^1$}
\date{\emph{actffa.tex}, \today }
\address{$^1$ Laboratoire Kastler Brossel, Sorbonne Universit\'e, CNRS, ENS-PSL, Coll\`ege de France, Campus Pierre et Marie Curie, 4 place Jussieu, 75005 Paris, France}
\address{$^2$ Conservatoire National des Arts et M\'etiers, 292 rue Saint Martin, 75003 Paris, France}


\begin{abstract}
We evaluate the accuracy to be expected for the measurement of free fall acceleration of antihydrogen in the GBAR experiment, accounting for the recoil transferred in the photodetachment process. We show that the uncertainty on the measurement of gravity comes mainly from the initial velocity dispersion in the ion trap so that the photodetachment recoil is not the limiting factor to the precision as a naive analysis would suggest. This result will ease the constraints on the choice of the photodetachment laser parameters. 

\end{abstract}

\section{Introduction}

The asymmetry between matter and antimatter observed in the Universe is one of the fundamental problems of modern physics challenging the Standard Model \cite{Charlton2017}. Hints on this puzzle are looked for in experiments testing CPT symmetry between matter and antimatter as well as the interaction of antimatter with gravity \cite{Hori2013,Bertsche2015,Yamazaki2020}. 
CPT tests are quite precise \cite{Yamazaki2013,Safronova2018} but tests on antimatter gravity have currently a limited precision, with the sign of gravity acceleration not yet known experimentally \cite{Alpha2013}.

The Einstein Equivalence Principle, a cornerstone of General Relativity \cite{Will2018}, implies that all objects fall with the same acceleration in a given gravity field, say the field of the Earth. This Weak Equivalence Principle has been tested with high precision by experiments working on a variety of macroscopic and atomic objects \cite{Wagner2012,Touboul2017,Viswanathan2018,Asenbaum2020}, and ambitious projects are developed at new CERN facilities to produce low energy antihydrogen \cite{Maury2014} with the aim of measuring the free fall of neutral antihydrogen atoms \cite{Bertsche2018,Pagano2020,Mansoulie2019}. Among them, the GBAR experiment (\emph{Gravitational Behaviour of Antihydrogen at Rest}) aims at measuring $\gbar$, the gravity acceleration of antihydrogen in the Earth gravity field, by timing the free fall of ultracold $\Hbar$ atoms \cite{Mansoulie2019,Indelicato2014,Perez2015}. 

The principle of GBAR experiment is based upon an original idea of T.~H\"ansch and J.~Walz \cite{Walz2004}. Antihydrogen ions  $\Hbar ^+$ are cooled in an ion trap by using laser cooling techniques. A short laser pulse is applied to detach the excess positron forming a neutral antiatom, with the laser pulse marking the start of the free fall. At the end of this free fall, the antiatom annihilates on the walls of the free fall chamber producing secondary particles, the detection of which makes it possible to reconstruct the positions in time and space of the annihilation event. The free fall acceleration $\gbar$ is deduced from a statistical analysis of annihilation events.  For an accurate determination of $\gbar$, it is crucial to understand how the photodetachment process modifies the distribution of velocities and then affects the statistics of annihilation events.

In a naive estimation of the sensitivity of the experiment, using only linear variation analysis, the precision would be proportional to the dispersion of the vertical velocity distribution and the free fall height (with the experimental numbers given later in this paper, the dispersion of initial altitude has a much smaller effect than that of initial velocity on the timing analysis \cite{Dufour2014}). 
As a large part of the velocity dispersion is due to excess energy delivered by the photodetachment process, this naive analysis could lead to restrict the excess energy and consequently limit photodetachment efficiency. 
The main goal of this paper, taking into account the correlations between vertical and horizontal velocities induced by photodetachment, is to show that the precision is mainly limited by the initial velocity dispersion before the photodetachment process. 

In the following, we first discuss the initial distribution of velocities before the free fall (\S 2) and the distribution of annihilation events after the free fall (\S 3). We then present a Monte-Carlo simulation of the measurement of the free fall acceleration $\gbar$ using first a simple geometry where atoms fall down to a horizontal detection plane (\S 4). Repeating the analysis for the geometry of a cylindrical chamber, we then find the unexpected result that the precision can be improved by the presence of the ceiling that intercept some trajectories (\S 5). We finally sum up the results obtained in this paper and the perspectives it opens (\S 6).

\section{Velocity distribution before free fall}

\begin{figure}
\begin{center}
\includegraphics[width=.5\linewidth]{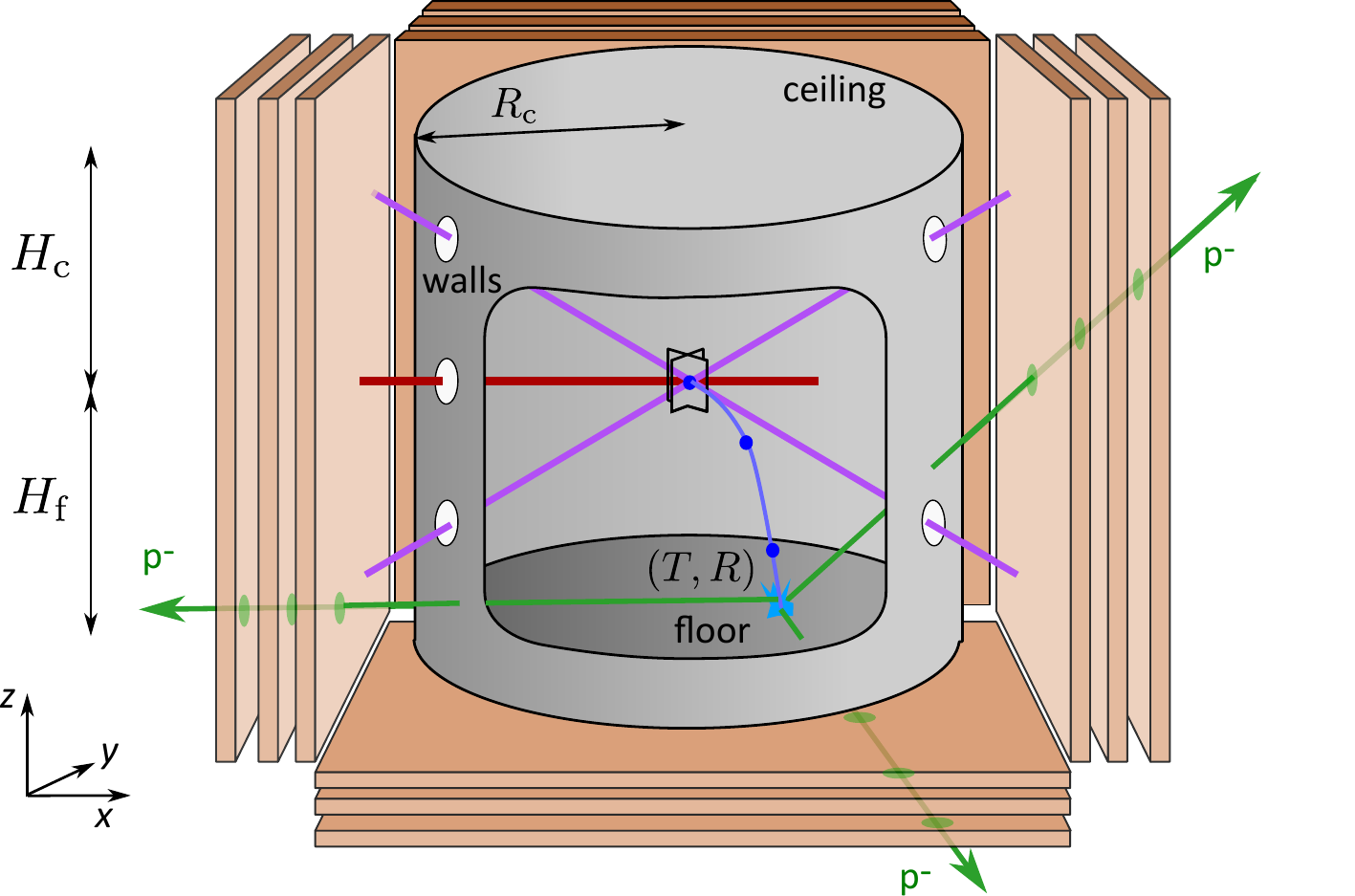}
\end{center}
\caption{Principle of the free fall measurement of antihydrogen in the GBAR experiment \cite{Perez2015}. $\bar{\mathrm H}^+$ ions are laser cooled in an ion trap. The excess positron is photo-detached by an horizontal laser (dark red line) to form a neutral $\bar{\mathrm H}$ atom that falls under gravity. Micromegas and scintillation detectors are used to reconstruct the positions in time and space of the annihilation events on the surfaces of the chamber. The dimensions of the cylindrical chamber are the free fall height $H_\mathrm{f}$, the height of the ceiling $H_\mathrm{c}$ above the trap and the radius $R_\mathrm{c}$ of the chamber.  }
\label{fig:FreeFallChamber}
\end{figure}

As already discussed, the principle of the experiment \cite{Walz2004} is to produce ultracold $\Hbar$ atoms from antihydrogen ions $\Hbar ^+$ sympathetically cooled in an ion trap \cite{Hilico2014,Sillitoe2017}. The acceleration $\gbar$ of $\Hbar$ atoms in the Earth's gravity field, simply called $g$ from now on, is measured by timing the free fall from the photodetachment of the excess positron to the annihilation of antiatoms when they touch the surfaces of the free fall chamber (see Fig.\ref{fig:FreeFallChamber}). 
Our analysis is focused on the optimization of uncertainty, expected to reach a value of the order of 1\% after analysis of the free fall of approximately 1000 atoms \cite{Indelicato2014,Perez2015}. All numerical studies will be done with the standard value $g_0=\SI{9.81}{\meter\second}^{-2}$.

We do not discuss the details of reconstruction of trajectories from the analysis of the impact of secondary particles on Micromegas and scintillation detectors surrounding the experiment chamber \cite{Radics2019} and assume that this process gives access to the time and space positions of the  annihilation event. We focus our attention on the influence of the initial velocity distribution on the measurement of free fall acceleration. We now discuss this distribution before and after the photodetachment process.

Before the photodetachment, the ion is prepared in the ground state of an ion trap, that we assume for simplicity to be harmonic and isotropic.  The initial wave packet has a minimal dispersion identical along the three space directions ($m$ the atomic mass, $\omega =2\pi f$ with $f$ the trap
frequency and $v=\frac{p}{m}$ the atomic velocity) 
\begin{eqnarray}
\Delta x=\Delta y=\Delta z=\zeta 
=\sqrt{\frac{\hbar }{2m\omega }}\quad, \quad
\Delta v_{x}=\Delta v_{y}=\Delta v_{z}=\Delta v
=\frac{\hbar }{2m\zeta }~. 
\label{eq:deltav}
\end{eqnarray}
Typical values of the velocity dispersion $\Delta v$ range from $\SI{0.077}{\meter\second}^{-1}$ to $\SI{0.45}{\meter\second}^{-1}$ for frequencies $f$ from $\SI{30}{\kilo\hertz}$ to $\SI{1}{\mega\hertz}$.

The ground state is represented by a wave-packet centred at the origin of space coordinates with a Gaussian shape in position and momentum representations
\begin{eqnarray}
&&\psi _{\mathrm trap}(\bm r_{\mathrm trap})
=\left( \frac{1}{2\pi \zeta ^{2}}\right) ^{3/4}\exp
\left( -\frac{\bm r_{\mathrm trap}^{2}}{4\zeta ^{2}}\right) ~, \\
&&\widetilde{\psi}_{\mathrm trap}(\bm p_{\mathrm trap})= \left( \frac{1}{2\pi \Delta p ^{2}}\right) ^{3/4}\exp
\left( -\frac{\bm p_{\mathrm trap}^{2}}{4\Delta p ^{2}}\right) \quad,\quad \Delta p\equiv m \Delta v~,
\label{eq:harmonic_trap_momentum}
\end{eqnarray}
where $\bm r_{\mathrm trap}$ and $\bm p_{\mathrm trap}$ are the position with respect to the trap centre and the momentum. This wavepacket is conveniently represented by a Wigner distribution \cite{Dufour2014,Wigner1932} which is a factorized quasi-probability distribution in phase space for a gaussian wave-packet, while fully describing quantum properties of the initial state.
We do not give more equations but emphasize that the photodetachment-induced recoil and the free fall evolution can both be described by classical laws in this Wigner  representation.


The photodetachment of the excess positron results from the absorption by the ion of a photon with an energy slightly higher than the threshold energy. For hydrogen, the threshold energy (electron affinity) is around $\SI{0.754}{\electronvolt}$ corresponding to a photon wavelength of $\SI{1.64}{\micro\meter}$. 
The photodetachment cross-section  depends on the excess energy $\delta E$ above the threshold \cite{Lykke1991,Vandevraye2014,Bresteau2017}. In the case of $\mathrm{H}^{-}$ or $\Hbar ^{+}$ atom, an s-electron is detached into an outgoing p-wave electron so that the cross-section scales as $\delta E^{\frac 32}$.  Photodetachment efficiency thus requires large enough value of the excess energy $\delta E$, but this leads to significant kinetic energy considered as a problem for the uncertainty in the naive linear variation analysis sketched in the Introduction.

\begin{figure}
\centering
\includegraphics[width=15cm]{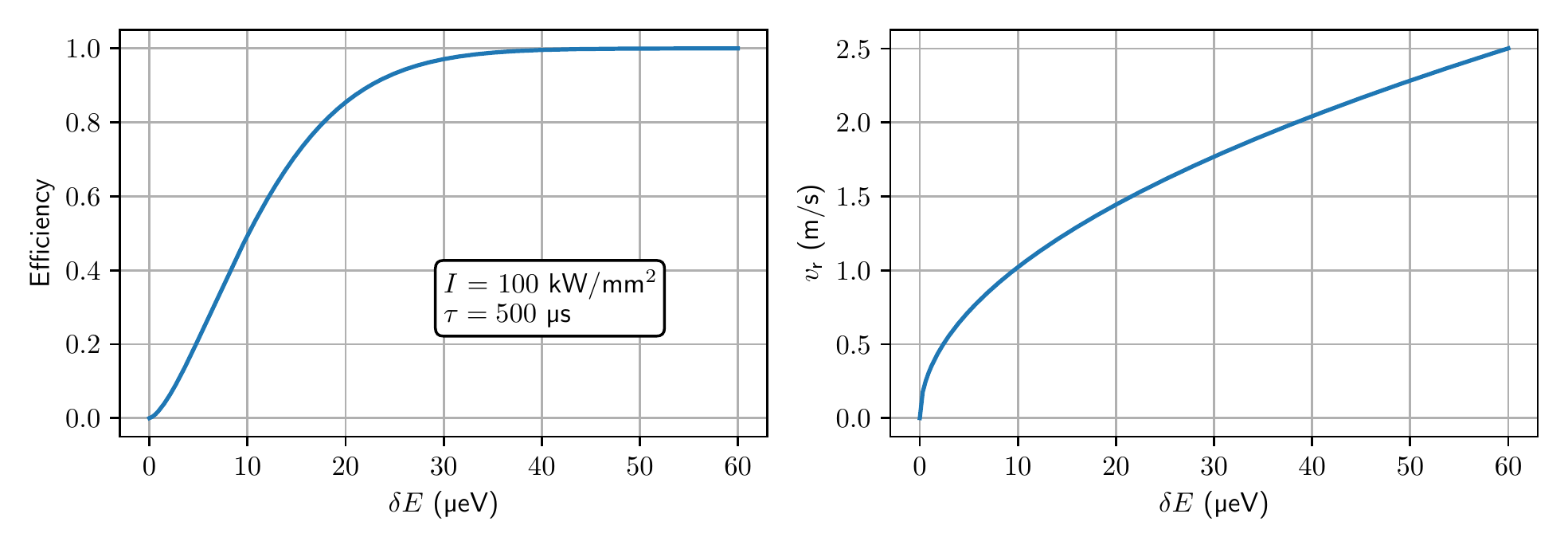}
\caption{Photodetachment efficiency and recoil velocity magnitude for the atom as a function of the excess energy $\delta E$ above the photodetachment threshold. Laser intensity is $\SI{100}{\kilo\watt/\milli\meter^2}$ and pulse duration $\SI{500}{\micro\second}$.}
\label{fig:photodetachment_efficiency}
\end{figure}

The recoil of the atom is calculated from energy and momentum conservation. We use here the approximation where it is described as an instantaneous elastic process, not affecting the position. The momentum transfer $\bm{q}$ is the sum of the photon momentum $\hbar \bm{k}$ and the opposite $\bm{q}_\mathrm{r}$ of the momentum of detached excess positron  (see Fig.~\ref{fig:photodetachment_efficiency})
\begin{equation}
\bm{q}=\hbar \bm{k}+\bm{q}_\mathrm{r}~.
\end{equation}
The photon recoil has a direction given by the laser propagation along $x-$axis and a magnitude fixed by spectroscopy
\begin{equation}
\frac{\hbar k}{m}\simeq \SI{0.24}{\meter\second^{-1}}~.
\label{eq:recoilphoton}
\end{equation}%
The recoil $\bm{q}_\mathrm{r}$ associated to excess positron has a fixed modulus $q_\mathrm{r}$ which in the limit $m_{e}\ll m$ ($m_{e}$ positron's mass) is given by 
\begin{equation}
\bm{q}_\mathrm{r}^{2}=q_\mathrm{r}^{2}\equiv 2m_{e}~\delta E \quad , \quad 
v_\mathrm{r}=\frac{q_\mathrm{r}}m=\frac{\sqrt{2m_{e}~\delta E}}m~.
\label{eq:recoilv}
\end{equation}
Typical values of the photodetachment recoil velocity $v_\mathrm{r}$ range from $\SI{1.25}{\meter\second}^{-1}$ to $\SI{2.50}{\meter\second}^{-1}$ for excess energies 
$\delta E$ from $\SI{15}{\micro\electronvolt}$ to $\SI{60}{\micro\electronvolt}$.
With these values of the excess energy, needed to reach a good photodetachment efficiency, $v_\mathrm{r}$ is neatly larger than the velocity dispersion $\Delta v$ before photodetachment, which would be a worrying problem within the naive precision analysis.

Fortunately, one can do better than predicted by the naive precision analysis, because the different components of the recoil velocity are correlated. Precisely, the recoil momentum associated to detached positron lies on a sphere with a center $\hbar \bm{k}$ and a radius $q_\mathrm{r}$.
The angular distribution on the sphere depends on the polarisation of the laser. 
In spherical coordinates, with poles given by the polarisation direction $\widehat{\mathbf{n}}$ of the photodetachment laser, the angular distribution probability is given by
\begin{equation}
\frac{\mathrm d\mathcal{P}}{\mathrm d\Omega} = \frac3{4\pi} \left(\frac{\bm q_\mathrm{r} \cdot \bm \widehat{\mathbf{n}}}{q_\mathrm{r}}\right)^2  ~.
\label{eq:angulardistrib}
\end{equation}

Though photodetachment is modelled by an instantaneous process, its precise time of occurence $t_0$ is not certain. The probability distribution of this time depends on the parameters of the laser (laser profile, excess energy and pulse duration). We call $\tau$ the typical width of this probability and assume for simplicity that the distribution of $t_0$ is given by a logistic distribution 
\begin{equation}
\delta_\tau(t_0) = \frac1{4\tau}\frac1{\cosh^2\left(\frac {t_0}{2\tau}\right)}
\label{eq:logistic}
\end{equation}

After the photodetachment, the momentum of the atom is given by the sum of the momentum inside the trap (distribution given by the wave function in momentum space \eqref{eq:harmonic_trap_momentum}) and the recoil momentum $q$ transfered in the photodetachment 
while the position is unaffected in the assumption of an instantaneous process
\begin{equation}
\bm p_0 = \bm p_\mathrm{trap} + \bm q
\quad,\quad \bm r_0 = \bm r_\mathrm{trap}~.
\end{equation}
The distribution of momenta $\bm p_0$ before the free fall therefore results from the convolution of the Gaussian distribution in the trap and the distribution of photodetachment recoil. 
Thereafter, we use velocity $\displaystyle{\bm v_0 = \frac{\bm p_0}m}$ instead of momentum as for free fall, the mass of the atom is not relevant any more.

\begin{figure}
\centering\includegraphics[width=\linewidth]{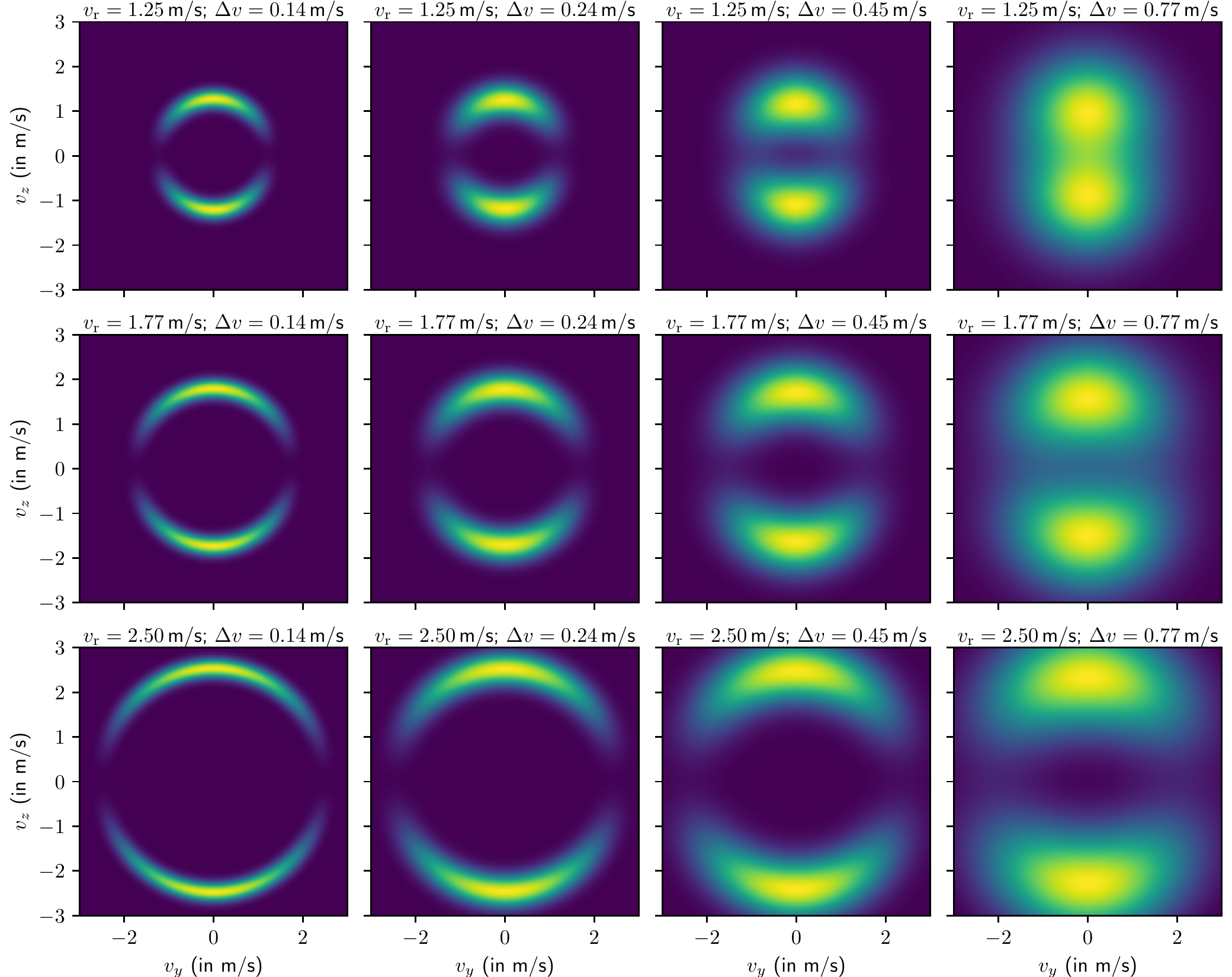}
\caption{Density plots of the velocity distributions in the $\left(v_{y},v_{z}\right) $ plane for 
$\delta E=\SI{15}{\micro\electronvolt}$, $\SI{30}{\micro\electronvolt}$ and \SI{60}{\micro\electronvolt} on the first, second and third lines respectively (corresponding to $v_\mathrm{r} = \SI{1.25}{\meter\second}^{-1}$, $\SI{1.77}{\meter\second}^{-1}$ and $\SI{2.50}{\meter\second}^{-1}$) and for $f=\SI{0.1}{\mega\hertz}$, $\SI{0.3}{\mega\hertz}$, $\SI{1}{\mega\hertz}$ and $\SI{3}{\mega\hertz}$
on the first, second, third and fourth columns respectively (corresponding to $\Delta v=\SI{0.14}{\meter\second}^{-1}$, $\SI{0.24}{\meter\second}^{-1}$, $\SI{0.45}{\meter\second}^{-1}$ and $\SI{0.77}{\meter\second}^{-1}$).}
\label{fig:momentumdistribution}
\end{figure}

The convoluted distributions $\Pi$ of velocities $\bm v_0$ can be calculated analytically. Its angular part depends only on the cosine $\displaystyle{\cos\theta = \frac{\bm v_0 \cdot \widehat{\mathbf{n}}}{v_0}}$ of the angle $\theta$ between the polarisation $\widehat{\mathbf{n}}$ and velocity $\bm{v_0}$.
The result of the convolution is 
\begin{eqnarray}
\Pi(\bm v_0) = \frac 1{\sqrt{2\pi}\Delta v}\left( e^{- \frac{\left(v_0 - v_\mathrm{r}\right)^2}{2 \Delta v^{2}}} - e^{- \frac{\left(v_0 + v_\mathrm{r}\right)^2}{2 \Delta v^2}} \right)\frac 3{4\pi v_\mathrm{r}v_0}\mathfrak{F}(\theta)~,
\label{eq:v0Distributio}   \\
\mathfrak{F}(\theta) \equiv \left( 1-2\mathfrak{F}_{\ast}\right)\cos^{2}\theta +\mathfrak{F}_{\ast}\sin^{2}\theta ~,~
\mathfrak{F}_{\ast} \equiv 
\frac{\varsigma^{2}}{\tanh{\left(\frac{1}{\varsigma^{2}} \right)}} - \varsigma^{4} ~ , ~ 
\varsigma^2 \equiv\frac{\Delta v^2}{v_{0} v_\mathrm{r}}~.
\notag
\end{eqnarray}
The distributions $\Pi \left(\bm v_0\right)$ are shown as density plots on Figure \ref{fig:momentumdistribution} for different values of the parameters $\delta E$ and $f$
(equivalently $v_\mathrm{r}$ and $\Delta v$). 
The distribution, invariant under a rotation around the polarization axis, is shown in the 
$\left(v_y,v_z\right) $ plane for a vertical polarization. 
Other polarizations are obtained by a 3D rotation of the distribution.

In most cases of experimental interest, the Gaussian velocity dispersion $\Delta v$ is smaller than the recoil velocity $v_\mathrm{r}$, and the distribution is restricted to the vicinity of the sphere of radius $v_\mathrm{r}$ with an angular density proportional to $\cos ^2\theta$. The full distribution, for arbitrary values of $\Delta v^2$, describes a gaussian smearing along radial variations and a gaussian smearing of the initial angular distribution.

\section{Distribution of annihilation events}

In this section, we get the distribution of annihilation events by using the fact that the Wigner representation connects points in phase space through classical laws for an Hamiltonian at most quadratic in the coordinates, which is the case for free fall in a constant gravity field. Hence classical free fall trajectories $\displaystyle{\bm r_t =\bm r_0 + \bm v_0 t + \bm g \frac{t^2}2}$ ($\bm r_0$ and $\bm v_0$ initial position and velocity and $\bm r_t$ position at time $t$) are used to connect the initial and final points in phase space.


The Monte-Carlo simulation of initial velocities is obtained by sampling random variables corresponding to the different sources of uncertainty. The distribution in the trap is generated using a normal distribution function with a dispersion \eqref{eq:deltav}. Such distribution is directly computed using available random generator. 

The recoil \eqref{eq:recoilphoton} induced by the absorption of the photon is constant and its direction is given by the laser direction, assumed to be along the $x-$axis. 
The recoil velocity associated to the excess positron $\bm{v}_\mathrm{r}$ has a fixed norm $v_\mathrm{r}$ given by the conservation of momentum and energy in eq.\eqref{eq:recoilv}.
In spherical coordinates, assuming that the polarisation is vertical, the probability of the velocity direction 
($\theta$, $\phi$) is obtained using eq.\eqref{eq:angulardistrib}.
This probability distribution is generated using two independent random variables $X_1$ and $X_2$ with uniform distributions in $[0, 1[$ producing random angles
\begin{equation}
    \Phi  = 2\pi X_1    \quad,\quad
    \Theta =\arccos{\sqrt[3]{1-2X_2}}  ~.
\end{equation}
If the polarisation forms an angle $\theta_n$ with the vertical, a rotation is performed to obtain
the velocity components. 
In order to sample the distribution of the photodetachment time $t_0$, a third random variable $X_3$ with uniform distributions in $[0, 1[$ is calculated and a time obeying the logistic distribution (\ref{eq:logistic}) is deduced as 
$\displaystyle{\tau \log(X_3/(1-X_3))}$.

The annihilation position and time are random variables that depend on the random initial velocity $\bm{v}_0$ and the random time of photodetachment $t_0$. For each initial velocity we calculate the trajectory of the atom. For simple surfaces like a plane or a vertical cylinder, the intersection point is calculated analytically. Geometries of chamber considered in this paper are composed of such surfaces. Impact parameters are computed for each surface and the first impact gives the annihilation event coordinates $\bm R$ and $t$. 

\begin{figure}
\begin{center}
\includegraphics[width=\linewidth]{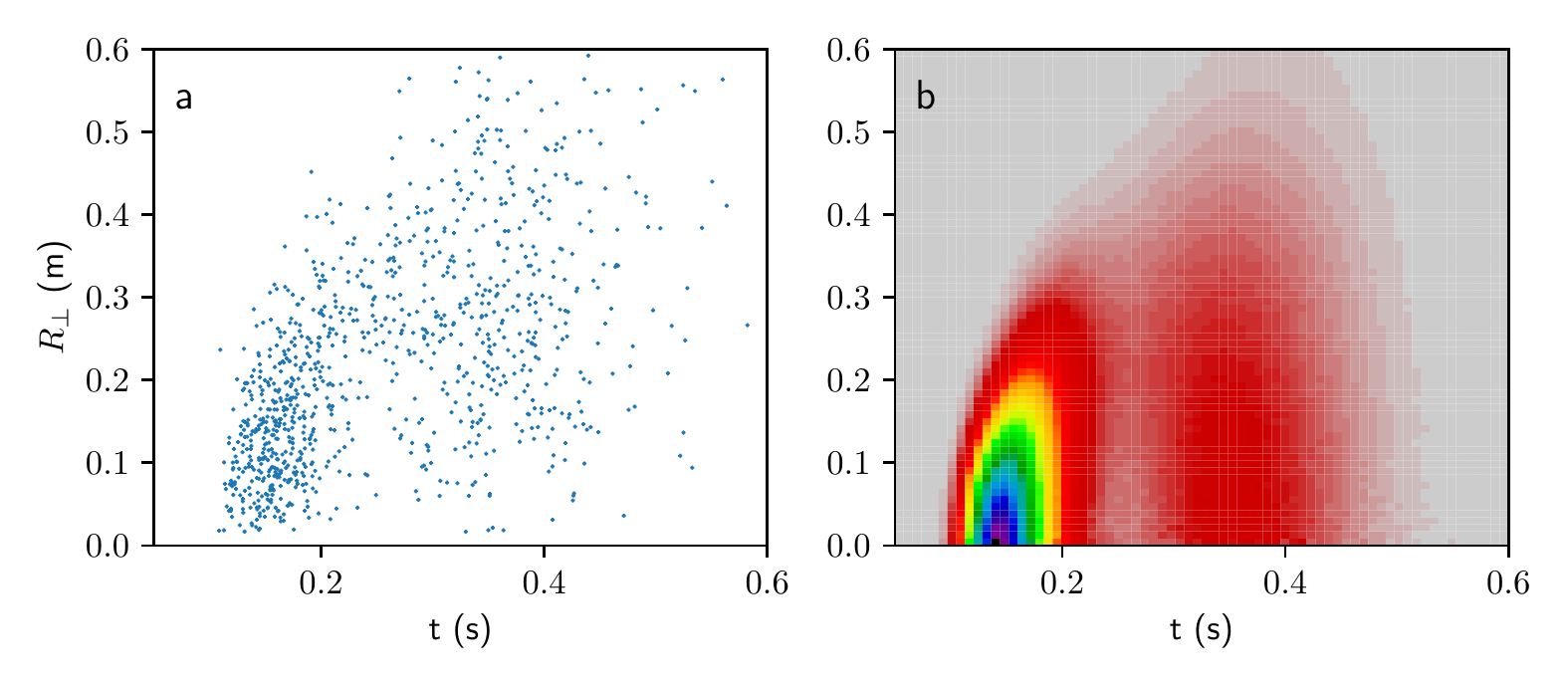}
\end{center}
\caption{
Monte-Carlo simulation of positions in space and time of the annihilation event on a horizontal plane placed at $\SI{30}{cm}$  below the trap (time $t$ and distance to the center $R_\perp = \sqrt{X^2 + Y^2}$) for $v_\mathrm{r}=\SI{1.25}{\meter\second}^{-1}$, $\Delta v=\SI{0.45}{\meter\second}^{-1}$ and vertical polarisation. 
On the right, 2D histogram calculated with \num{10000000} points. Each point is weighted by $1/R_\perp$ in order to take into account the surface on the chamber associated with each square of the 2D histogram.}
\label{fig:MC_and_J}
\end{figure}

\begin{figure}
\begin{center}
\includegraphics[width=.7\linewidth]{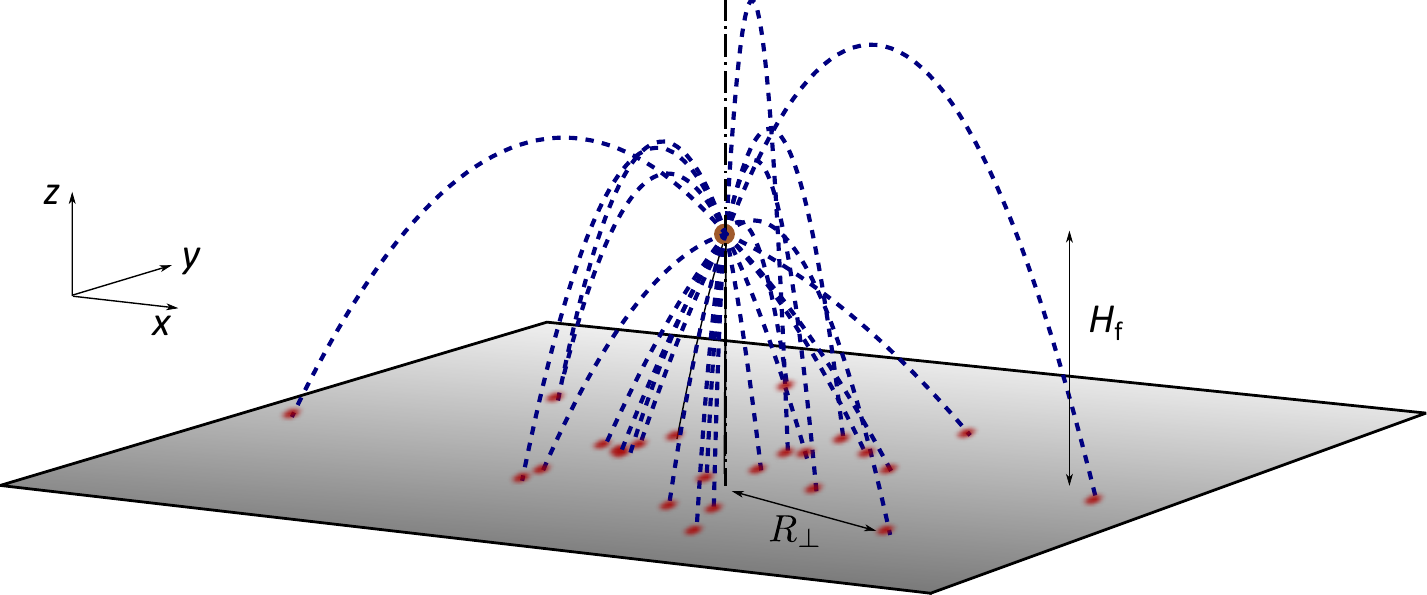}
\end{center}
\caption{Simple geometry for the GBAR experiment~: free fall on a horizontal plane placed at a distance $H_\mathrm f$ below the trap.}
\label{fig:schema_simplified}
\end{figure}

Results of typical Monte-Carlo simulations are depicted on Fig.~\ref{fig:MC_and_J}. We have plotted a scatter of $1000$ points and a 2D histogram of the annihilation position and time obtained with \num{10000000} points. Atoms are detected on the floor and a vertical polarisation is assumed. For a given position, the distribution of annihilated events presents two peaks corresponding to atoms with different initial velocities annihilated at the same point with different time of flight. Due to the correlation between the vertical and horizontal velocity induced by the photodetachment, the position of the peaks depends on the distance $R_\perp$ to the center as illustrated on figure \ref{fig:schema_simplified}. 

The 2D histogram corresponds to the probability, per unit of time and per unit of surface, to hit the chamber at a given position and time. In order to have this interpretation, the weight of each point in graph was set to $1/R_\perp$ so that the surface on the chamber associated with a square of the histogram is constant. This quantity correspond to the local flux of atoms. In the next section, in order to analyse the Monte-Carlo simulation and estimate $g$ from a draw of annihilation events, we will derive an analytical expression of the probability current.


We denote $\bm R$ and $T$ the position in space and time of an annihilation event
\begin{equation}
T = t_0 + t \quad,\quad
\bm R = \bm r_0 + \bm r = \bm r_0 + \bm v_0 t + \bm g  \frac{t^2}2 ~,
\end{equation}
where $t_0$ is the photodetachment time (within the distribution of width $\tau$), $t$ the time of flight 
and $\bm r$ the displacement during this time. 
We also introduce the velocity $\bm v$
\begin{equation}
\bm v = \bm v_0 + \bm g t ~.
\end{equation}
The quantities with lower case $\bm r$, $t$ correspond to the displacement and delay during free fall while the quantities with upper case $\bm R$, $T$ correspond to observed position and time of detection. As already discussed, the initial dispersion in position plays a negligible role in the problem, whereas the uncertainty on the photodetachment time $t_0$ has to be accounted for.

From values of $\bm r$ and $t$, one can deduce the initial velocity $\bm v_0$ and final velocity $\bm v$:
\begin{eqnarray}
\bm v_0 = \frac{\bm r}{t} - \bm g \frac{t}2 \quad,\quad
\bm v = \frac{\bm r}{t} + \bm g \frac{t}2 ~.
\end{eqnarray}
Because there is a single initial velocity corresponding 
to a position in space and time, the probability density
$\rho(\bm r, t)$ to find a particule at position  $\bm r$ is proportional to velocity distribution $\Pi(\bm v_0)$ before the free fall. We obtain the detection probability current $\bm j(\bm r, t) = \bm v \rho(\bm r, t)$ as follows
\begin{equation}
\bm j(\bm r, t) = \frac{\Pi(\bm v_0)}{t^3}\left(\frac{\bm r}{t} + \bm g \frac{t}2\right) ~.
\label{eq:current}
\end{equation}
Accounting for the dispersion of $t_0$, the observed current of $\bm R$ and $T$ is finally given by
\begin{equation}
\bm J(\bm R, T) = \int  \bm j(\bm R - \bm r_0, T - t_0) \delta_\tau(t_0)\rho_0(\bm r_0) \mathrm d^3\bm r_0  \mathrm dt_0\simeq\int  \bm j(\bm R , T - t_0) \delta_\tau(t_0) \mathrm dt_0~,
\label{eq:Current}
\end{equation}
where we used the fact that initial position dispersion $\zeta$ has a negligible effect.

\begin{figure}
\centering
\withlabel[.2, -.2]{a}{\includegraphics[height=5.7cm]{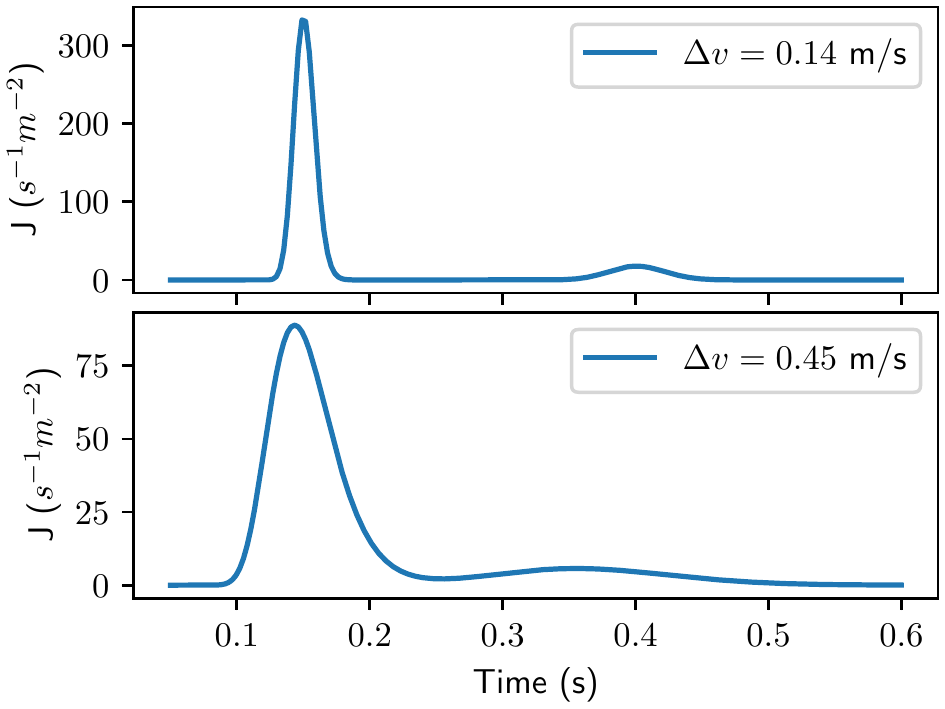}}\hfill
\withlabel[.2, -.2]{b}{\includegraphics[height=5.7cm]{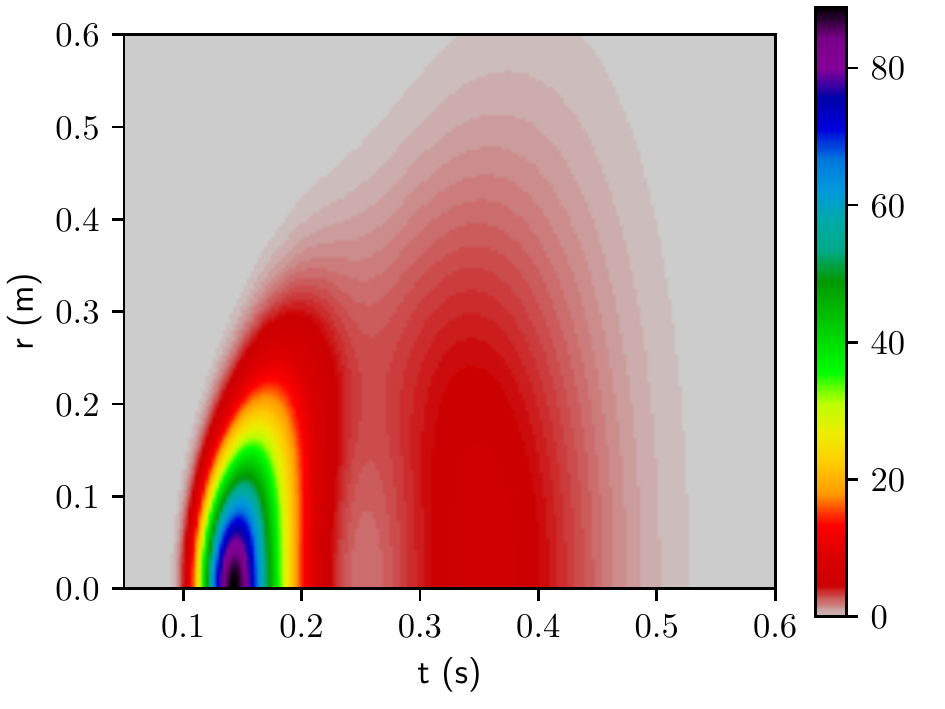}}
\caption{\textbf{a} - Distribution of annihilation times, at the center of the floor, for a vertical polarization
and two sets of parameters~: 
$\Delta v=\SI{0.14}{\meter\second}^{-1}$ and $v_\mathrm{r}=\SI{1.25}{\meter\second}^{-1}$ on the upper plot, $\Delta v=\SI{0.45}{\meter\second}^{-1}$ and $v_\mathrm{r}=\SI{1.25}{\meter\second}^{-1}$ on the lower one.
\textbf{b} - Distribution of annihilation events ($J(\bm R, T)$ in \si{s^{-1}m^{-2}}) on the floor as a function of $T$ and distance from the vertical axis, for $\Delta v=\SI{0.45}{\meter\second}^{-1}$ and $v_\mathrm{r}=\SI{1.25}{\meter\second}^{-1}$
}
\label{fig:figure_J_R_T}
\end{figure}

Atoms are detected when they reach the surface of the vacuum chamber. The observable quantity is the current through the surface, written at a given position $\bm R$ with $\bm u$ the unit vector orthogonal to the surface
\begin{eqnarray}
J(\bm R, T) &=& \bm u \cdot \bm J(\bm R, T) ~.
\end{eqnarray}
The quantity $J(\bm R, T)$ is the probability per unit of surface and per unit of time to detect a particle at position $\bm R$ and time $T$. 
Examples of such distributions are shown on figure \ref{fig:figure_J_R_T}a for a detection at the center of the floor. The plots are drawn for a vertical polarization and two sets of parameters $\Delta v=\SI{0.14}{\meter\second}^{-1}$ and $v_\mathrm{r}=\SI{1.25}{\meter\second}^{-1}$ on the upper plot, $\Delta v=\SI{0.45}{\meter\second}^{-1}$ and $v_\mathrm{r}=\SI{1.25}{\meter\second}^{-1}$ on the lower one.

The distributions presents two peaks corresponding to atoms with different initial velocities which are annihilated at the same point but different time of flight. The heights and shapes of these two peaks are depending on the precise values of the parameters entering the expression of $J\left( \bm{R},T\right)$.

The distribution of annihilation current is also represented on Fig~\ref{fig:figure_J_R_T}b by using false colour picture allowing to see both time and spatial coordinates. Due to the correlation between the vertical and horizontal velocity induced by the photodetachment, the maxima depends on the distance $R_\perp$ to the center (they do not form vertical lines) showing the importance to take into account the position is the analysis.

\section{Accuracy in the measurement of $g$}

In this section, we analyse the accuracy to be
expected in the experiment. Our main tool is the numerical Monte-Carlo method which closely mimics the data analysis to be developed for the experiment. We also present an analytical Cramer-Rao method, and use the good agreement of the two methods as a cross validation of the results.

\subsection{Monte-Carlo simulation}

From a draw of $N$ positions $\bm R_i$ and times $T_i$ of the annihilation events, one calculates the likelihood function $\mathcal L$ and the normalized likelihood function $\ell$
\begin{equation}
\mathcal L(g) = \prod_{i=0}^{N-1} J_g({\bm R_i}, T_i) \quad,
\quad
\ell(g) = \frac{\mathcal L(g)}{\int \mathcal L(g) \mathrm dg} ~.
\end{equation}

\begin{figure}
\centering
\withlabel[1.2, -.5]{a}{\includegraphics[width=0.48\linewidth]{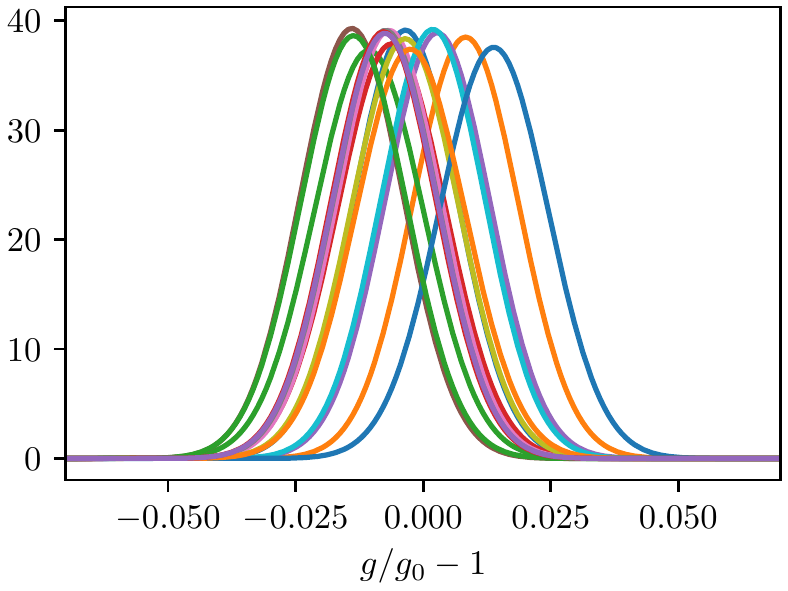}}%
\withlabel[1.2, -.5]{b}{\includegraphics[width=0.48\linewidth]{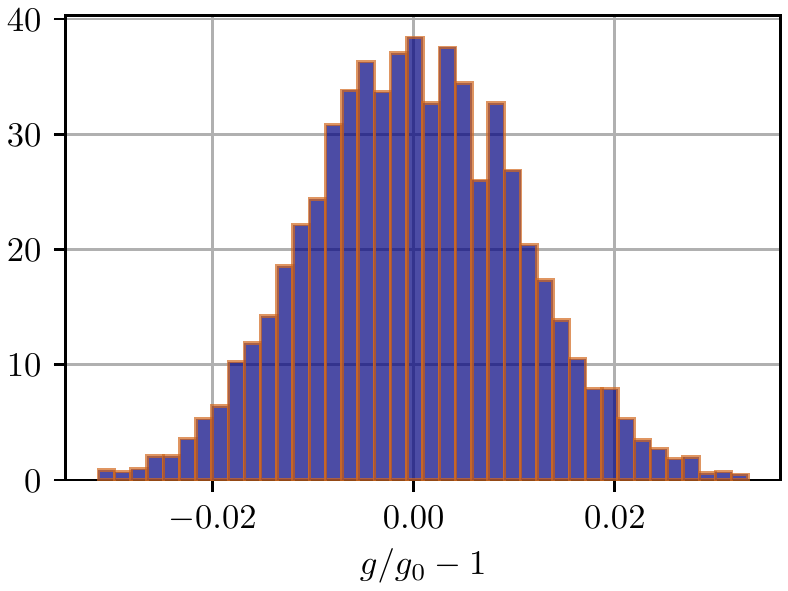}}
\caption{\textbf{a} - Normalized likelihood functions 
obtained for 15 independent random draws of 1000 atoms.
\textbf{b} - Normalized histogram of the estimators $\widehat{g}$ for 5000 random draws, for the geometry depicted on Figure~\ref{fig:schema_simplified}}.
\label{fig:randomdraws}
\end{figure}

Normalised likelihood functions $\ell(g)$ are plotted on  Figure~\ref{fig:randomdraws}a, for independent random draws, each one with 1000 atoms.
The estimator $\widehat{g}$ is obtained numerically by maximising the likelihood function $\mathcal L(g)$ or $\ell(g)$. We have plotted an histogram of such estimators $\widehat{g}$ for $M=5000$ independent random draws on  Fig.~\ref{fig:randomdraws}b. An accuracy $\sigma_{g}^\mathrm{MC}$ is obtained by calculating the standard deviation in this histogram. 

The reduced likelihood functions have Gaussian shapes with close values of their variance, but they are shifted with
respect to one another, hence giving different values of the estimator. Each likelihood function thus leads to an estimation of the variance close to the variance deduced
from the histogram of values on a large number of independent
random draws. These observations indicate that the statistical efficiency of the analysis is close to $1$, as confirmed by the Cramer-Rao analysis discussed below.


In the Cramer-Rao method, the dispersion $\sigma_{g}^\mathrm{CR}$ of the estimated $\widehat{g}$ 
is deduced from the Fisher information $\mathcal{I}_{g}$
calculated from the $g-$dependence of the distribution \cite{Frechet,Cramer,Refregier} ($\mathbb{E}$ denotes an
expectation value, $J_{g}$ the current probability calculated for a trial value $g$ and $N$ the number of events) 
\begin{eqnarray}
&&\sigma_{g}^\mathrm{CR}=\sqrt{\frac{1}{N\mathcal{I}_{g}}}~, \\
\mathcal{I}_{g} &=&\mathbb{E}\left[ -\frac{\partial ^{2}}{\partial g^{2}}
\ln J_{g}\right] =\mathbb{E}\left[ \left( \frac{\partial }{\partial g}\ln
J_{g}\right) ^{2}\right] =
\int \text{d}S\text{d}T~
\frac{\left( \partial _{g}J_{g}\right) ^{2}}{J_{g}}~. \end{eqnarray}
The last integral is taken over the surface of the chamber and annihilation time.
With the analytical expressions given above, the evaluation of
$\sigma_{g}^\mathrm{CR}$ is reduced to the numerical integration of the Fisher information $\mathcal{I}_{g}$.

The Cramer-Rao dispersion $\sigma_{g}^\mathrm{CR}$ 
corresponds to an optimal estimation of the parameter $g$.
Dispersions $\sigma_{g}^\mathrm{MC}$ obtained in Monte-Carlo simulations are thus expected to be slightly larger than  $\sigma_{g}^\mathrm{CR}$, the difference being small for a good statistical efficiency \cite{Cramer}. We have checked that this is indeed the case for all dispersions we have calculated, taking into account the uncertainties in Monte-Carlo estimations obtained with a finite numbers of events
(a few examples are discussed below).
This confirms the good statistical efficiency, which was expected for an experiment with 1000 atoms and a smooth probability distribution. From an experimental point of view, a good efficiency means that the unique random draw to be obtained in the experiment is representative of the variety of
results for different random draws in the numerical simulations.

\subsection{Discussion of the results}

We now discuss variations of the obtained accuracy versus parameters, using Monte-Carlo or Cramer-Rao methods equivalently as they lead to the same conclusions. We consider first a simple geometry where annihilation of antihydrogen takes place only on a horizontal plane placed below the trap at an altitude $H_f$ (in practice, this corresponds to results obtained by pushing the ceiling at a high enough altitude and the walls at a large enough radius so that all annihilations occur on the floor).

\begin{figure}
\centering
{\includegraphics[width=.70\linewidth]{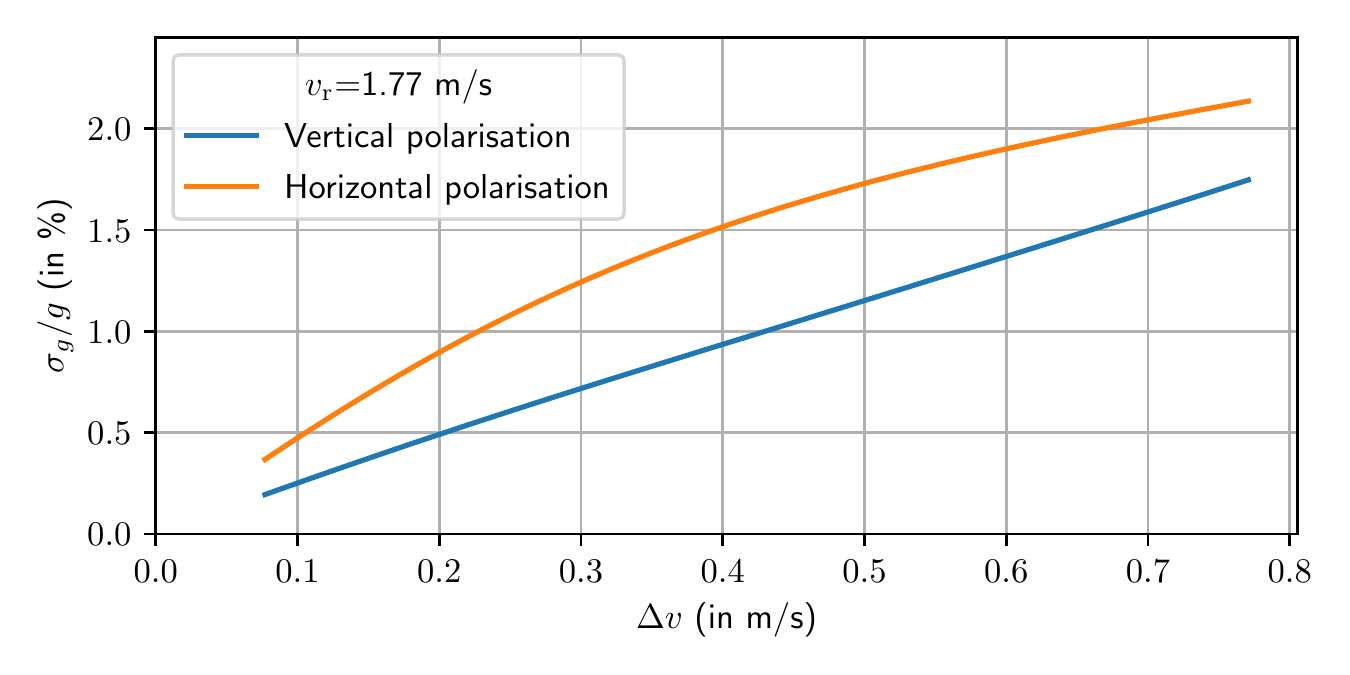}}
\caption{Variations of predicted relative dispersion, as a function of velocity dispersion $\Delta v$ for vertical and horizontal polarisation and a recoil velocity $v_\mathrm{r}=\SI{1.77}{\meter\second}^{-1}$ 
($\delta E=\SI{30}{\micro\electronvolt}$). }
\label{fig:variationsfrequency}
\end{figure}

Figure \ref{fig:variationsfrequency} shows that the accuracy is improved (the relative dispersion is decreased) for lower values of $\Delta v$, the two curves corresponding to vertical and horizontal polarizations but the same excess energy.
Meanwhile, the accuracy is also improved for a vertical polarization, which stands in contrast with the expectation of the naive linear variation analysis sketched in the Introduction. The variance of vertical recoil is indeed smaller for an horizontal polarisation than for a vertical one, so that naive expectations would lead to a larger dispersion in the former case than in the latter one. 
With the results of the full analysis performed in this paper, the accuracy is better for vertical polarization. This can be traced to the fact that atoms with an initial upwards velocity have a longer time of flight, which is beneficial for the determination of $g$.

\begin{figure}[t]
\centering
{\includegraphics[width=.70\linewidth]{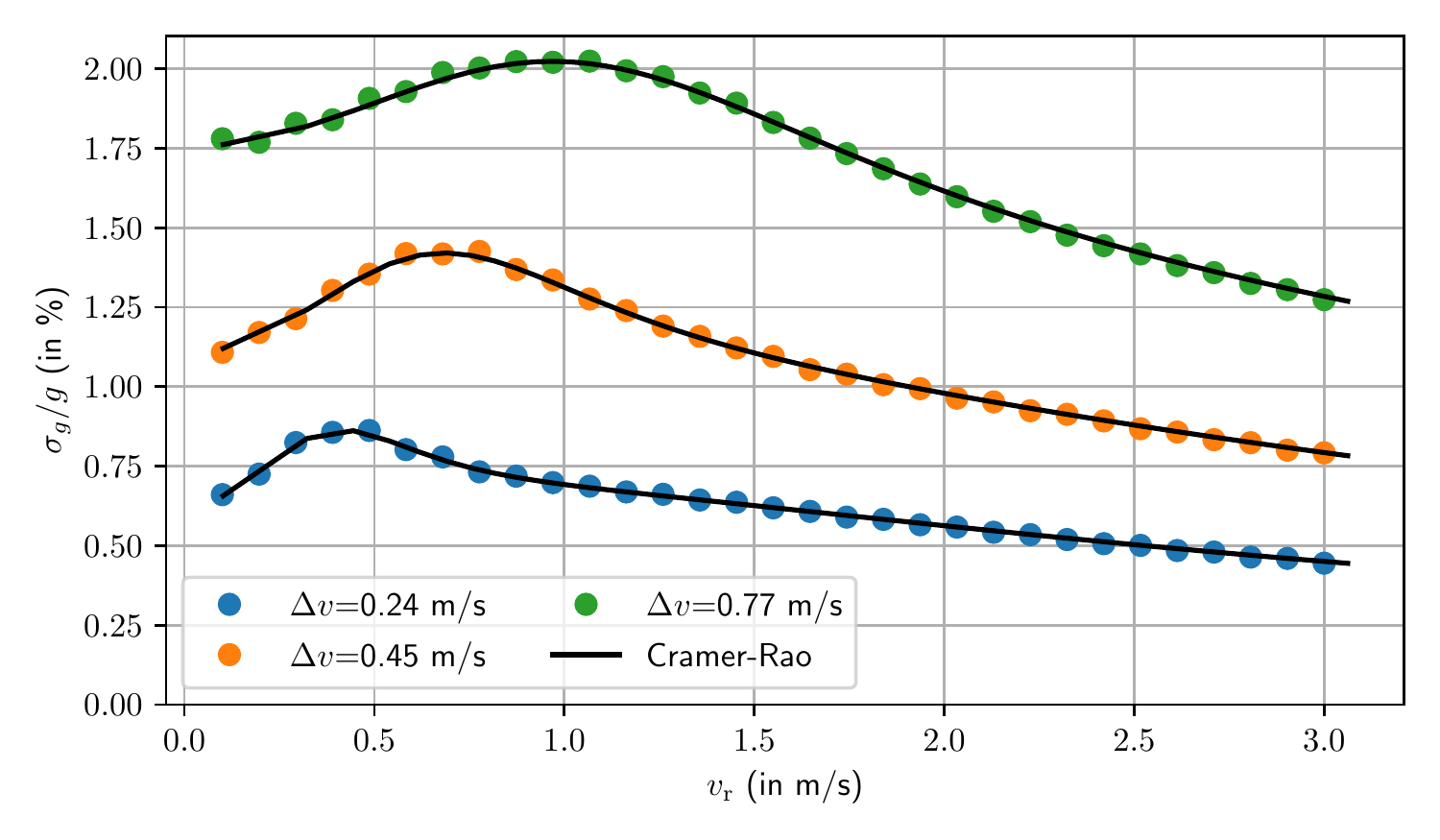}}
\caption{Variations of predicted relative dispersions, as a function of recoil velocity $v_\mathrm{r}$, for different velocity dispersion $\Delta v$ and a vertical polarization. Full lines represent Cramer-Rao predictions
while dots show the results of Monte-Carlo simulations.}
\label{fig:variationsfrequencypolfreq}
\end{figure}

The effect of the photodetachment recoil is shown on Fig.\ref{fig:variationsfrequencypolfreq} with full lines representing Cramer-Rao predictions and dots showing the results of Monte-Carlo simulations.
One clearly sees on the plot that the results of the two methods are close, which confirms the good statistical efficiency.
One also notices a conclusion standing in contrast with the prediction of a dispersion proportional to the initial velocity dispersion, since the accuracy is improved for larger values of the excess energy. 

\begin{figure}[b]
\centering{\includegraphics[scale=0.8]{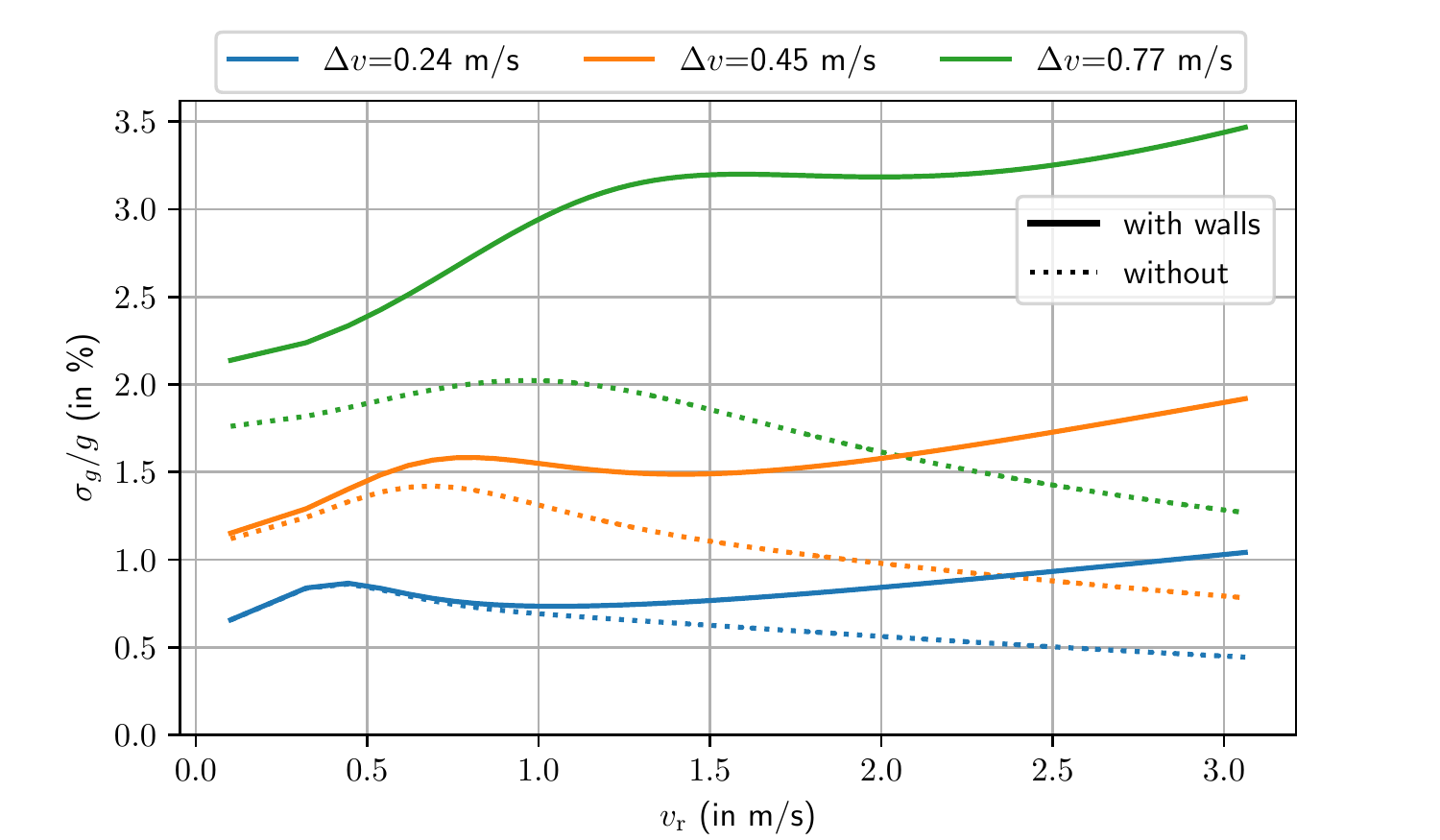}}
\caption{Variations of predicted relative dispersions, as a function of recoil velocity $v_\mathrm{r}$, for different velocity dispersion $\Delta v$ and a vertical polarization. Full lines represent Cramer-Rao predictions with the walls at a radius $R_\mathrm{c}=$\SI{30}{\centi\meter} and dashed lines the predictions with the walls pushed to a much larger radius.}
\label{fig:WithorWithoutWalls}
\end{figure}

Up to now, we have considered only free fall on the horizontal floor. However, the experiment will take place in a chamber with finite radius and it is necessary to take into account the presence of the walls on which atoms having large initial velocities will annihilate. 
Figure~\ref{fig:WithorWithoutWalls} shows the comparison of the relative uncertainty for the configuration without walls and with walls on a cylinder of radius $R_\mathrm{c}=\SI{30}{\centi\meter}$. For very low values of $v_\mathrm{r}$ and $\Delta v$, the results are only slightly modified as most atoms still annihilate on the floor. 
For high values of $v_\mathrm{r}$ and $\Delta v$ in contrast, a significant fraction of atoms reach the wall before the floor, and their time of flight is reduced as well as the effective free fall height, leading to a detrimental effect on the accuracy which is clearly seen on the curves.

\subsection{Conclusions at this point}

A crucial conclusion of the results presented up to now is that the accuracy is determined mainly by the Gaussian velocity distribution in the ion trap. The dispersion in the photodetachment recoil velocity does not add noise in the data analysis even when it is larger than that of initial Gaussian velocities. This is due to the fact that the vertical and horizontal components of recoil velocity are correlated (they lie on a sphere), a property which has been used in the analysis to mitigate their contribution to uncertainty analysis.

The too naive linear variation analysis sketched in the Introduction leads to the prediction of a dispersion proportional to the dispersion of initial vertical velocity. It gives in most cases an accuracy poorer than the correct results produced by the analysis in this paper. It also favours the choice of an horizontal polarization whereas the correct results lead to prefer a vertical polarization. The results obtained here thus reduce the constraints on the choice of photodetachment parameters, as they allow an increase of $\delta E$ which is certainly good for discussion of the photodetachment efficiency.

\section{Full geometry of the experimental chamber}

In this section, we study the geometry of the free fall
chamber designed for the GBAR experiment \cite{Perez2015}, with a cylindrical chamber having finite dimensions for the free fall height $H_\mathrm{f}$, the height of the ceiling $H_\mathrm{c}$ above the trap and the radius $R_\mathrm{c}$ of the cylinder (see Fig.\ref{fig:FreeFallChamber}). In order to specify numbers, we choose here $H_\mathrm{f}=H_\mathrm{c}=\SI{30}{\centi\meter}$ and $R_\mathrm{c}=\SI{25}{\centi\meter}$ (calculations can be done as well by changing these dimensions).

Annihilations can thus take place on the floor, the walls or the ceiling, with proportions depending on the parameters of
the velocity distribution before free fall. For simplicity, we consider that the annihilation occurs on the first material surface touched by freely falling antiatom, thus disregarding the possibility of quantum reflection on the surfaces \cite{Dufour2013} (calculations accounting for quantum reflection will be presented in a forthcoming paper). Even with these simplifications, there appear new problems to be studied, coming from the presence of steps in the detection current on the floor or walls, due to prior annihilation on the ceiling.

\begin{figure}
\withlabel[1.1, -.3]{a}{\includegraphics[width=0.48\linewidth]{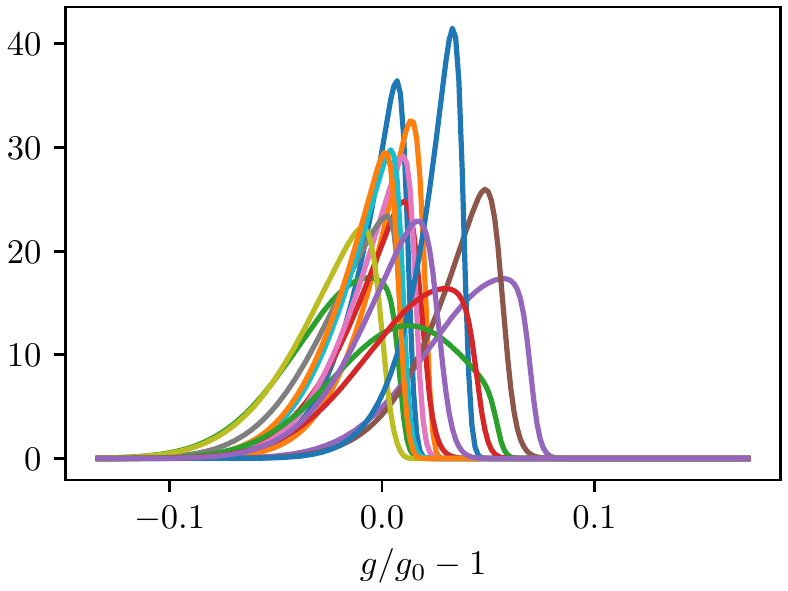}} %
\withlabel[1.1, -.3]{b}{\includegraphics[width=0.48\linewidth]{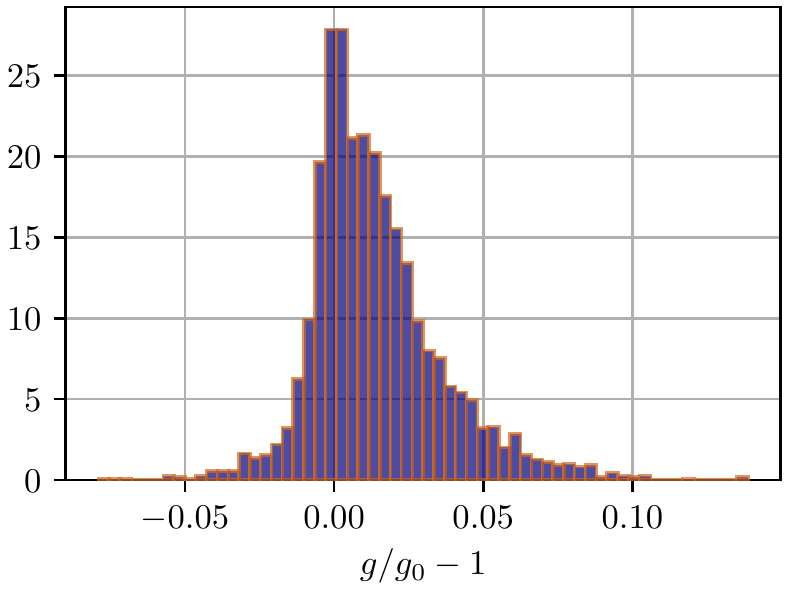}}
\caption{\textbf a: Normalized likelihood functions 
obtained random draws of 1000 atoms.
\textbf b: Normalized histogram of the estimators 
$\widehat{g}$ for 3000 random draws simulated with $v_\mathrm r = \SI{2.5}{m/s}$, $\Delta_v =\SI{0.77}{m/s}$ and a vertical polarization.}
\label{fig:randomdraws_cover}
\end{figure}

We apply the same procedures as before to perform Monte-Carlo simulations and deduce maximum likelihood estimators $\widehat{g}$. The results, depicted on Figure~\ref{fig:randomdraws_cover}, show striking differences with Figure~\ref{fig:randomdraws}. The likelihood functions calculated for one random draw of $N=1000$ atoms are no longer Gaussian while the histogram of estimated $\widehat{g}$ for a number of different simulations has an asymmetric shape with a significant bias. This behaviour, appearing when we include the ceiling in the geometry, can be qualitatively understood by analyzing the current  $J\left( \bm{R},T\right)$.

\begin{figure}
\withlabel[.2, -.5]{a}{\includegraphics[height=5.5cm]{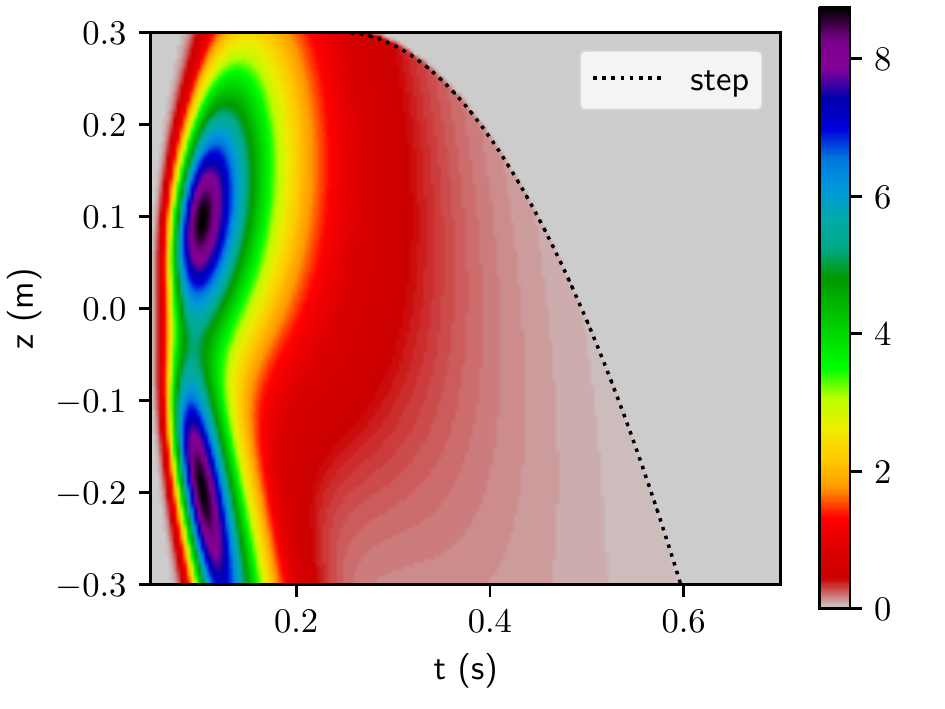}}
\withlabel[.2, -.5]{b}{\includegraphics[height=5.5cm]{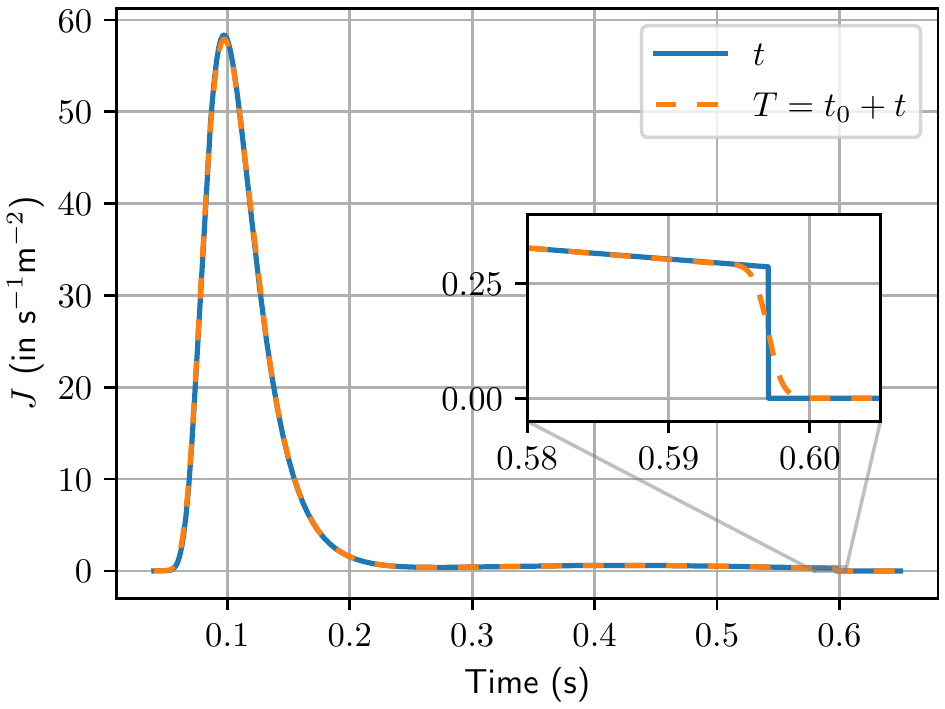}}
\caption{$\textbf{a}$ : Distribution of annihilation events ($J(\bm R, T)$ in \si{s^{-1}m^{-2}}) on the wall
as a function of $T$ and altitude $z$ with the presence of the ceiling $H_\mathrm c=\SI{30}{cm}$, for $\Delta v=\SI{0.77}{\meter\second}^{-1}$, $v_\mathrm{r}=\SI{2.17}{\meter\second}^{-1}$ and vertical polarisation. $\textbf{b}$ : Comparison between $j(\bm R, t)$ (blue line) and $J(\bm R, T)$ (orange dash line) for $\bm R$ at the bottom of the chamber, for $\Delta v=\SI{0.77}{\meter\second}^{-1}$, $v_\mathrm{r}=\SI{2.17}{\meter\second}^{-1}$ and vertical polarisation.}
\label{fig:figure_J_z_T_obstacles}
\end{figure}

We have represented on Figure~\ref{fig:figure_J_z_T_obstacles}-a the particle current on the wall as a function of the height $z$ and time of flight $t$ with a ceiling at distance $H_\mathrm c=\SI{30}{cm}$ above the trap. Due to the ceiling, a shadow appears in the graph as there is a critical time of flight $t_c$ above which atoms reach the ceiling before the wall, for a given position $z$
\begin{equation}
  t_c(z) =\sqrt{\frac{2H_\mathrm c}g} 
  + \sqrt{\frac{2(H_\mathrm c-z)}g} ~. 
\end{equation}
Above this critical time, the current would be strictly 0 if the start time was perfectly known (case $\tau=0$).
The Figure~\ref{fig:figure_J_z_T_obstacles}-b, represents the current on the floor as a function of the time of flight (drawn for a finite $\tau$). The critical time appears as a step on the current, smoothed by taking into account the uncertainty in the photodetachment time (dashed orange line).
For example, the critical time is \SI{597}{\milli\second} on the floor $z=-H_\mathrm f$ for the default parameters. 

\begin{figure}
\withlabel[.2, -.5]{a}{\includegraphics[width=0.49\linewidth]{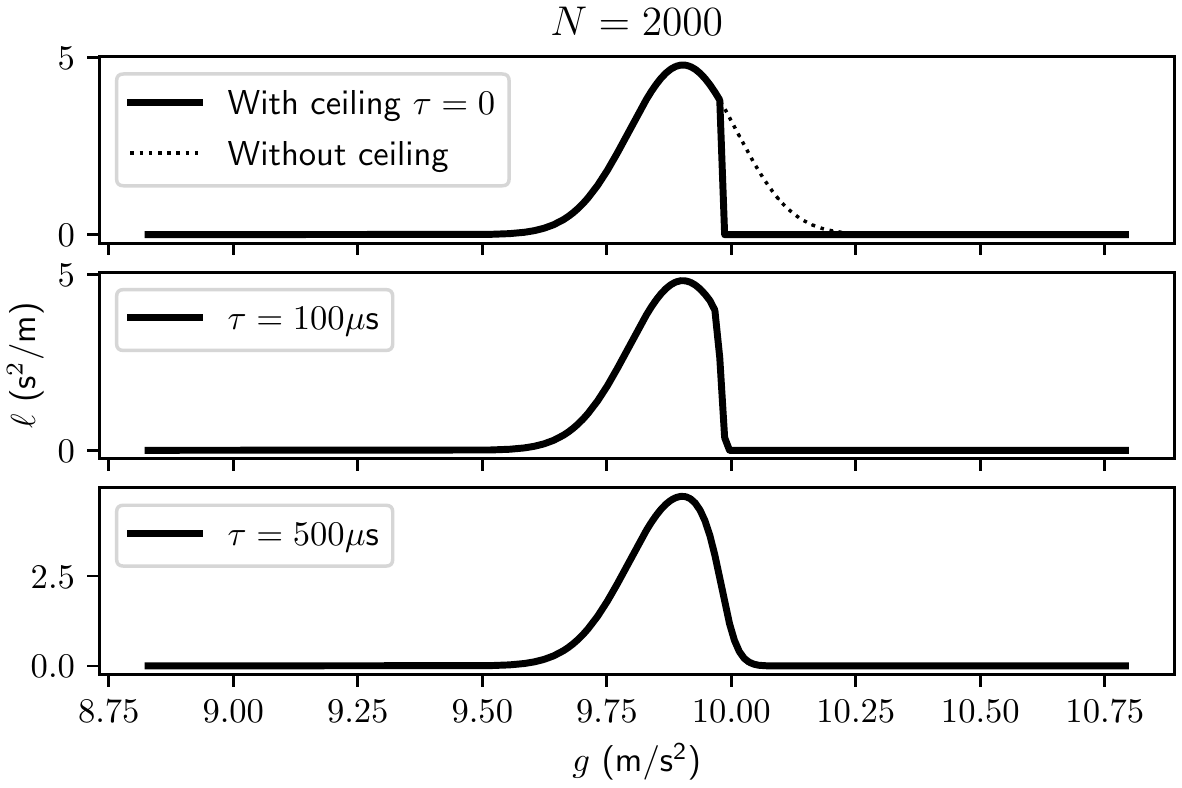}} %
\withlabel[.2, -.5]{b}{\includegraphics[width=0.49\linewidth]{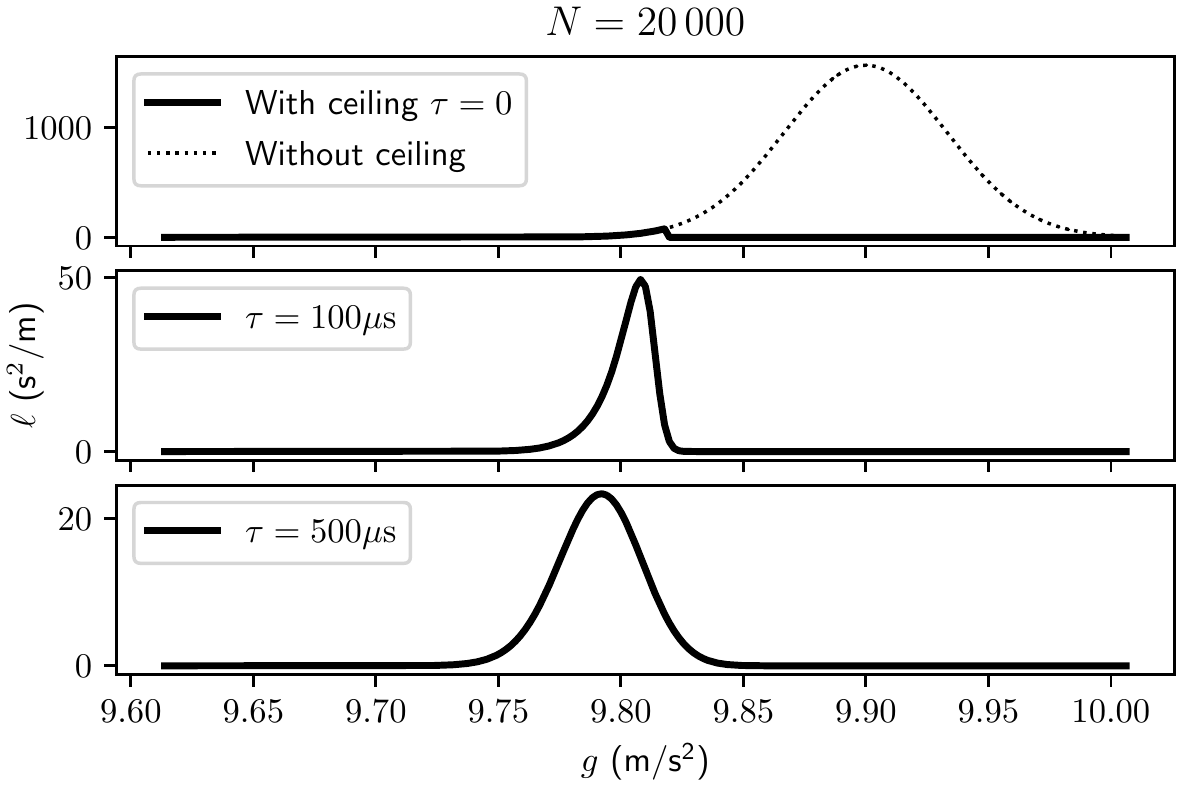}}
\caption{Normalized likelihoods for a set of $N=2000$ and $N=20000$ points. The continuous lines represent normalized likelihood for different values of $\tau$ and dashed lines likelihood with the ceiling pushed up to a higher height to calculate the current on the wall and floor (normalized using the same factor as for $\tau=0$)}.
\label{fig:likelihood_cover}
\end{figure}

The effect of the shadow and the presence of this critical time is visible in the likelihood function. We have represented on the Fig.~\ref{fig:likelihood_cover} the likelihood for a set of 2000 and 20000 draws and for different values of $\tau$. With the ceiling pushed up to a higher height (dashed line), we observe that the likelihood is a Gaussian, as expected from the large number of sample. This Gaussian is biased towards large $g$ as, on average, atoms arrive on the wall or floor with a shorter time of flight than without. If we take into account the presence of the ceiling for the calculation of the current on the wall, we observe a sudden drop in the likelihood. This drop corresponds to the first value of $g$ such that there is an atom in the shadow created by the presence of the ceiling. This sudden drop is smoothed by the dispersion of the photodetachment time. 

\begin{figure}
\centering
\withlabel[.2, -.5]{}{\includegraphics[width=0.6\linewidth]{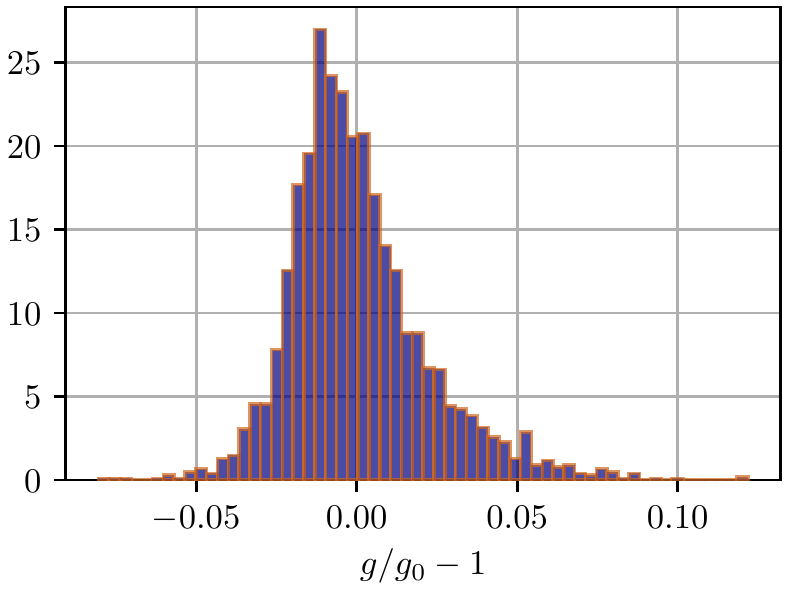}} %
\caption{Normalized histogram of the estimators $\check{g}$ for 3000 random draws simulated with $v_\mathrm r = \SI{2.5}{m/s}$, $\Delta_v =\SI{0.77}{m/s}$ and a vertical polarization.}
\label{fig:mean_likelihood}
\end{figure}

If the position of the cut in the current occurs for a value of $g$ smaller than the maximum of the Gaussian calculated without the ceiling (dashed line Fig.~\ref{fig:likelihood_cover}), the maximum likelihood is given by the position of the cut. The latter is determined by a single event (the first point that enters into the shadow of the ceiling when $g$ is changed) and explains the bias towards high value of $g$ observed in Fig.~\ref{fig:randomdraws_cover}b. In order to reduce this bias, we define a new estimator $\check g$ as the mean value of the normalised likelihood $\ell$ 
\begin{equation}
\check g = \int g\ell(g) \mathrm dg = \frac{\int g\mathcal L(g) \mathrm dg}{\int \mathcal L(g) \mathrm dg} ~.
\end{equation}
An histogram of the values of $\check g$ is shown on Fig.~\ref{fig:mean_likelihood}. The parameters are the same as on Fig.~\ref{fig:randomdraws_cover} b. The uncertainty of $\check g$ is similar to the uncertainty of $\widehat{g}$, but the relative bias is reduced to a small value \num{0.05}\% (instead of \num{1.5}\% for $\widehat{g}$). The precise value of the bias could be different for another simulation (it also shows fluctuations) but it would in any case remain much smaller than the dispersion.

\begin{figure}\centering
\includegraphics[width=0.48\linewidth]{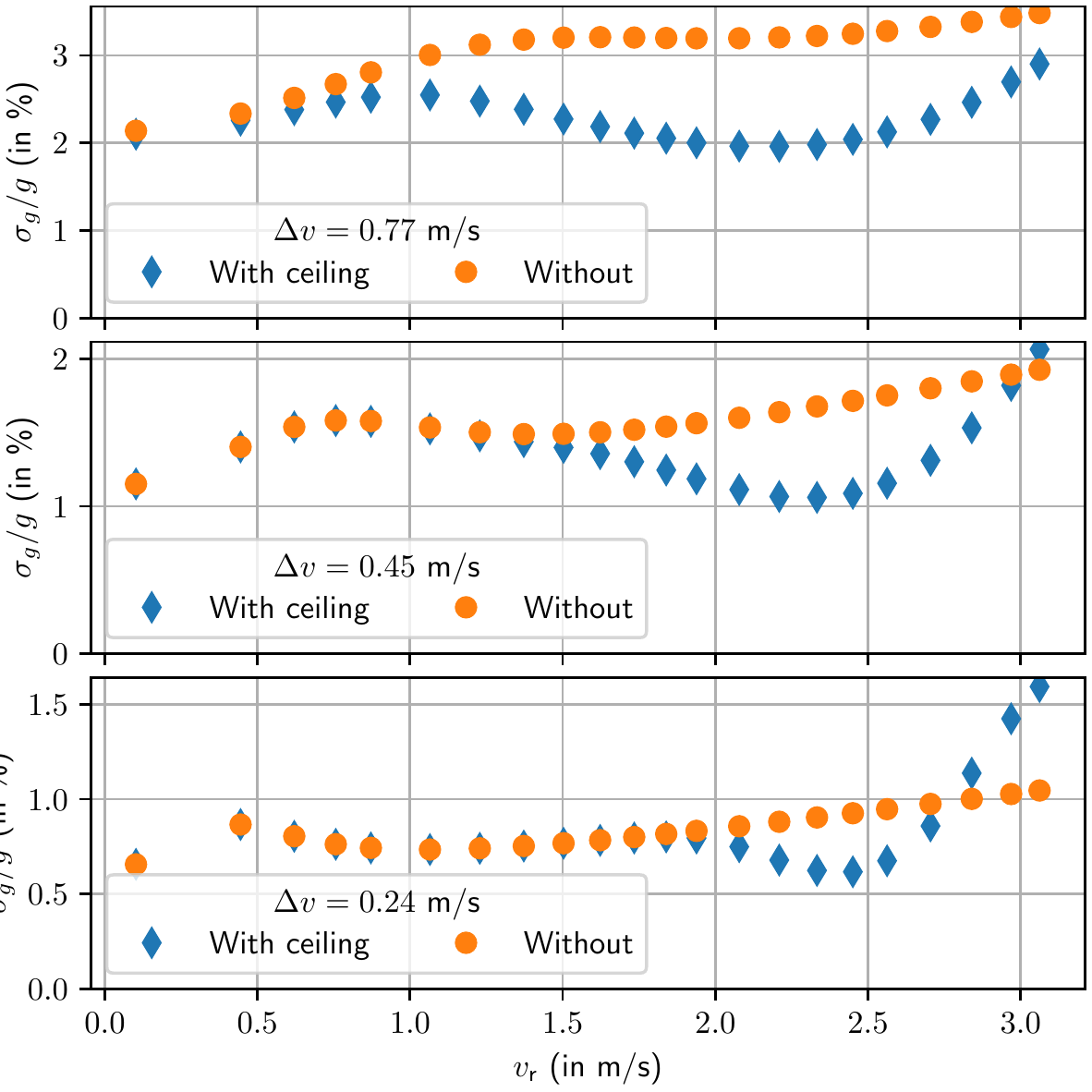} %
\caption{ Uncertainty on the estimation of $g$ as a function of recoil velocity $v_\mathrm{r}$ and for different values of the initial velocity dispersion $\Delta v$. Each point correspond to a Monte-Carlo simulation with $N=1000$ atoms. Orange circles correspond to simulations without the ceiling, blue diamonds to simulations with the ceiling.}
\label{fig:improvement_due_to_ceiling}
\end{figure}

Using the estimator $\check g$, we calculate the uncertainty for various parameters (see Fig.~\ref{fig:improvement_due_to_ceiling}) and clearly see that the ceiling can improve it. For an initial velocity dispersion $\Delta v=\SI{0.45}{\meter\second}^{-1}$ and vertical polarisation, we reach an optimum of 1.1\% with $v_\mathrm{r}=\SI{2.33}{\meter\second}^{-1}$ ($\delta E = \SI{52}{\micro\electronvolt}$). The optimum is at 2.0\% with $v_\mathrm{r}=\SI{2.17}{\meter\second}^{-1}$ ($\delta E = \SI{45}{\micro\electronvolt}$) for $\Delta v=\SI{0.77}{\meter\second}^{-1}$. 

\begin{figure}[b!]
\centering
\withlabel[.5, -.5]{\footnotesize a}{\includegraphics[width=0.47\linewidth]{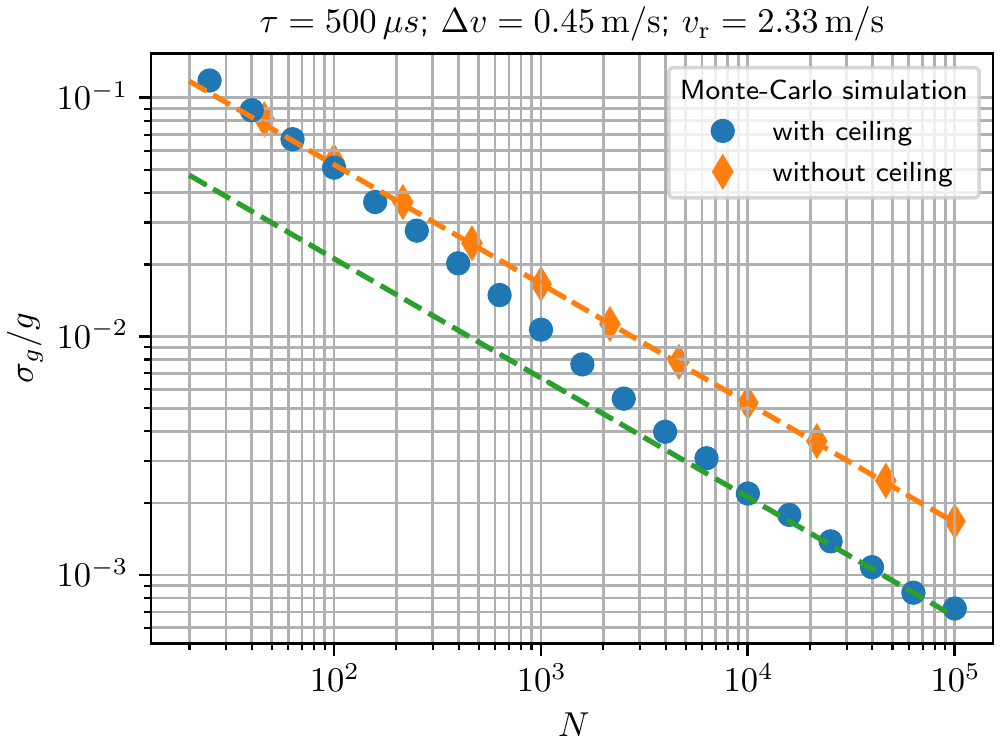}}
\withlabel[.5, -.5]{\footnotesize b}{\includegraphics[width=0.47\linewidth]{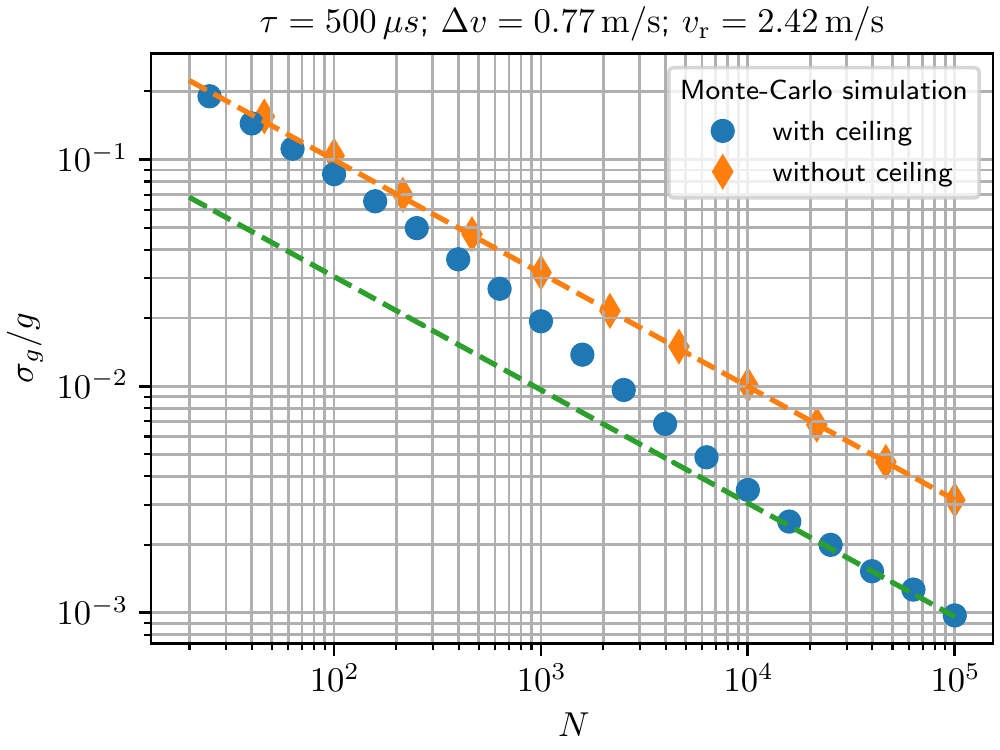}} 
\caption{Uncertainty on the estimation of $g$ as a function of function of the number $N$ of initial event for $\tau=\SI{500}{\micro\second}$. The green line corresponds to the Cramer-Rao bound without ceiling and the orange one to the Cramer-Rao bound with ceiling using $\tau=\SI{500}{\micro\second}$. Both Cramer-Rao bound scale as $1/\sqrt{N}$. }
\label{fig:courbe_pierre_opt}
\end{figure}

On Fig.~\ref{fig:courbe_pierre_opt} we plot the uncertainty as a function of the number of initial events $N$, for three values of the pair of parameters ($v_\mathrm{r}$, $\tau$), and  observe three regimes. In the first regime, corresponding to small value of $N$ ($N\ll 100$), the uncertainty is mainly given by the Cramer-Rao bound which would be calculated without ceiling (in practice with the ceiling pushed to a high enough altitude above the trap), as the probability to have an event close to the critical line is thus negligible. In the opposite regime corresponding to a very large value of $N$ ($N\gg 10000$), many events are close to this line so that the uncertainty is given by the Cramer-Rao bound calculated with the ceiling in place. The precision of the measurement of $g$ thus depends on the resolution of this boundary, that is the value of $\tau$. In the intermediate regime ($100 \ll N \ll 10000$), the presence of the ceiling brings information but not enough to reach the good efficiency limit. The uncertainty is thus seen to decrease faster that $1/\sqrt{N}$ and it does not depend on the dispersion of the photodetachment time.

We thus understand quantitatively the behaviour of the uncertainty on Fig.~\ref{fig:improvement_due_to_ceiling}. The effect of  the ceiling is optimal when a large number of atoms have an initial velocity close to the critical initial velocity $v_{\mathrm c} = \sqrt{2gH_\mathrm c}$. For $H_\mathrm c=\SI{30}{\centi\meter}$, we obtain $v_{\mathrm c} = \SI{2.4}{\meter\per\second}$.
Contrary to what one might have thought, the presence of the ceiling can improve the measurement uncertainty when $N$ is large enough. The measurement of $g$ is thus mainly the measurement of the critical time $t_c$ with an uncertainty varying as $1/N$ as long as one is not limited by the dispersion $\tau$ of the photodetachment time. For very large values of $N$ one reaches the Cramer-Rao regime with a limit that scales as $\sqrt{\tau/N}$.

\section{Conclusion}

In this paper, we have evaluated the accuracy to be expected for the measurement of free fall acceleration $g$ of antihydrogen in the GBAR experiment, accounting for the recoil transferred in the photodetachment process. 

Using Monte-Carlo simulations and analytical calculations of the Cramer-Rao bound, we have shown that the final accuracy is mainly determined by the width of the initial velocity dispersion in the trap and not by the larger velocity dispersion due to the photodetachment process. This follows from the strong correlation between vertical and horizontal components of the photodetachment recoil velocity which mitigate their contribution to the final uncertainty. This important result will allow to set a large enough excess energy above photodetachment threshold (say $\delta E>\SI{30}{\micro\electronvolt}$), so that the photodetachment probability is much more favourable. 

This work opens the way to a more realistic description of the experiment, which should take into account details which have not been treated here. Some have been evoked above, such as the reconstruction of annihilation events from the detection of secondary particles \cite{Radics2019} or the possible quantum reflection of antiatoms reaching the surfaces \cite{Dufour2013}. Other ones involve the detailed design of the trap implying the presence of material surfaces in the vicinity of the trap \cite{Hilico2014}.

We have shown that the presence of the ceiling intercepting some trajectories of the antiatoms creates a shadow zone on the detection current which can be used to improve significantly the measurement accuracy. This counter-intuitive effect will be studied in greater details in a forthcoming publication, with the aim of better understanding the limits of the improvement in statistical analysis, and also of using this new effect in order to push the uncertainty to the best possible level. 

\subsection*{Acknowledgements}
We thank our colleagues in the GBAR collaboration \cite{GBARpage} for insightful discussions, in particular F. Biraben, P.P. Blumer, P. Crivelli, P. Debu, A. Douillet, N. Garroum, L. Hilico, P. Indelicato, G. Janka, J.-P. Karr, L. Liszkay, B. Mansouli\'e, V.V. Nesvizhevsky, F. Nez, N. Paul, P. P\'erez, C. Regenfus, F. Schmidt-Kaler, A.Yu. Voronin, S. Wolf.
This work was supported by the Programme National GRAM of CNRS/INSU with INP and IN2P3 co-funded by CNES.

\section*{References}

\bibliography{gbar.bib}

\providecommand{\newblock}{}
\begin{thebibliography}{10}
\expandafter\ifx\csname url\endcsname\relax
  \def\url#1{{\tt #1}}\fi
\expandafter\ifx\csname urlprefix\endcsname\relax\def\urlprefix{URL }\fi
\providecommand{\eprint}[2][]{\url{#2}}

\bibitem{Charlton2017}
Charlton M, Mills A~P and Yamazaki Y 2017 {\em Journal of Physics B: Atomic,
  Molecular and Optical Physics\/} {\bf 50} 140201
  \urlprefix\url{https://doi.org/10.1088/1361-6455/aa75d8}

\bibitem{Hori2013}
Hori M and Walz J 2013 {\em Progress in Particle and Nuclear Physics\/} {\bf
  72} 206--253
  \urlprefix\url{https://www.sciencedirect.com/science/article/pii/S0146641013000069}

\bibitem{Bertsche2015}
Bertsche W~A, Butler E, Charlton M and Madsen N 2015 {\em Journal of Physics B:
  Atomic, Molecular and Optical Physics\/} {\bf 48} 232001
  \urlprefix\url{https://doi.org/10.1088/0953-4075/48/23/232001}

\bibitem{Yamazaki2020}
Yamazaki Y 2020 {\em Proceedings of the Japan Academy, Series B\/} {\bf 96}
  471--501 \urlprefix\url{https://doi.org/10.2183/pjab.96.034}

\bibitem{Yamazaki2013}
Yamazaki Y and Ulmer S 2013 {\em Annalen der Physik\/} {\bf 525} 493--504
  \urlprefix\url{https://onlinelibrary.wiley.com/doi/abs/10.1002/andp.201300060}

\bibitem{Safronova2018}
Safronova M~S, Budker D, DeMille D, Kimball D~F~J, Derevianko A and Clark C~W
  2018 {\em Rev. Mod. Phys.\/} {\bf 90} 025008
  \urlprefix\url{https://link.aps.org/doi/10.1103/RevModPhys.90.025008}

\bibitem{Alpha2013}
{Alpha Collaboration}, {Amole} C, {Ashkezari} M~D, {Baquero-Ruiz} M, {Bertsche}
  W, {Butler} E, {Capra} A, {Cesar} C~L, {Charlton} M, {Eriksson} S, {Fajans}
  J, {Friesen} T, {Fujiwara} M~C, {Gill} D~R, {Gutierrez} A, {Hangst} J~S,
  {Hardy} W~N, {Hayden} M~E, {Isaac} C~A, {Jonsell} S, {Kurchaninov} L,
  {Little} A, {Madsen} N, {McKenna} J~T~K, {Menary} S, {Napoli} S~C, {Nolan} P,
  {Olin} A, {Pusa} P, {Rasmussen} C~{\O}, {Robicheaux} F, {Sarid} E, {Silveira}
  D~M, {So} C, {Thompson} R~I, {van der Werf} D~P, {Wurtele} J~S, {Zhmoginov}
  A~I and {Charman} A~E 2013 {\em Nature Communications\/} {\bf 4} 1785
  \urlprefix\url{https://www.nature.com/articles/ncomms2787}

\bibitem{Will2018}
Will C~M 2018 {\em Theory and Experiment in Gravitational Physics (new
  edition)\/} ({Cambridge University Press}) ISBN 9781107117440

\bibitem{Wagner2012}
Wagner T~A, Schlamminger S, Gundlach J~H and Adelberger E~G 2012 {\em Classical
  and Quantum Gravity\/} {\bf 29} 184002
  \urlprefix\url{https://doi.org/10.1088/0264-9381/29/18/184002}

\bibitem{Touboul2017}
Touboul P, M\'etris G, Rodrigues M, Andr\'e Y, Baghi Q, Berg\'e J, Boulanger D,
  Bremer S, Carle P, Chhun R, Christophe B, Cipolla V, Damour T, Danto P,
  Dittus H, Fayet P, Foulon B, Gageant C, Guidotti P~Y, Hagedorn D, Hardy E,
  Huynh P~A, Inchauspe H, Kayser P, Lala S, L\"ammerzahl C, Lebat V, Leseur P,
  Liorzou F, List M, L\"offler F, Panet I, Pouilloux B, Prieur P, Rebray A,
  Reynaud S, Rievers B, Robert A, Selig H, Serron L, Sumner T, Tanguy N and
  Visser P 2017 {\em Phys. Rev. Lett.\/} {\bf 119}(23) 231101
  \urlprefix\url{https://link.aps.org/doi/10.1103/PhysRevLett.119.231101}

\bibitem{Viswanathan2018}
Viswanathan V, Fienga A, Minazzoli O, Bernus L, Laskar J and Gastineau M 2018
  {\em Monthly Notices of the Royal Astronomical Society\/} {\bf 476}
  1877--1888 \urlprefix\url{https://doi.org/10.1093/mnras/sty096}

\bibitem{Asenbaum2020}
Asenbaum P, Overstreet C, Kim M, Curti J and Kasevich M~A 2020 {\em Phys. Rev.
  Lett.\/} {\bf 125}(19) 191101
  \urlprefix\url{https://link.aps.org/doi/10.1103/PhysRevLett.125.191101}

\bibitem{Maury2014}
Maury S, Oelert W, Bartmann W, Belochitskii P, Breuker H, Butin F, Carli C,
  Eriksson T, Pasinelli S and Tranquille G 2014 {\em Hyperfine Interactions\/}
  {\bf 229} 105--115 \urlprefix\url{https://doi.org/10.1007/s10751-014-1067-y}

\bibitem{Bertsche2018}
Bertsche W~A 2018 {\em Philosophical Transactions of the Royal Society A:
  Mathematical, Physical and Engineering Sciences\/} {\bf 376} 20170265
  \urlprefix\url{https://royalsocietypublishing.org/doi/abs/10.1098/rsta.2017.0265}

\bibitem{Pagano2020}
Pagano D, Aghion S, Amsler C, Bonomi G, Brusa R~S, Caccia M, Caravita R,
  Castelli F, Cerchiari G, Comparat D, Consolati G, Demetrio A, Noto L, Doser
  M, Evans A, Fani M, Ferragut R, Fesel J, Fontana A, Gerber S, Giammarchi M,
  Gligorova A, Guatieri F, Haider S, Hinterberger A, Holmestad H, Kellerbauer
  A, Khalidova O, Krasnick{\'{y}} D, Lagomarsino V, Lansonneur P, Lebrun P,
  Malbrunot C, Mariazzi S, Marton J, Matveev V, Mazzotta Z, M\"uller S~R,
  Nebbia G, Nedelec P, Oberthaler M, Pacifico N, Penasa L, Petracek V, Prelz F,
  Prevedelli M, Ravelli L, Rienaecker B, Robert J, R{\o}hne O, Rotondi A,
  Sandaker H, Santoro R, Smestad L, Sorrentino F, Testera G, Tietje I~C,
  Widmann E, Yzombard P, Zimmer C, Zmeskal J and Zurlo N 2020 {\em Journal of
  Physics: Conference Series\/} {\bf 1342} 012016
  \urlprefix\url{https://doi.org/10.1088/1742-6596/1342/1/012016}

\bibitem{Mansoulie2019}
Mansouli{\'e} B and {on behalf of the GBAR Collaboration} 2019 {\em Hyperfine
  Interactions\/} {\bf 240} 11
  \urlprefix\url{https://doi.org/10.1007/s10751-018-1550-y}

\bibitem{Indelicato2014}
Indelicato P, Chardin G, Grandemange P, Lunney D, Manea V, Badertscher A,
  Crivelli P, Curioni A, Marchionni A, Rossi B, Rubbia A, Nesvizhevsky V,
  {Brook-Roberge} D, Comini P, Debu P, Dupr{\'e} P, Liszkay L, Mansouli{\'e} B,
  P{\'e}rez P, Rey J~M, Reymond B, Ruiz N, Sacquin Y, Vallage B, Biraben F,
  Clad{\'e} P, Douillet A, Dufour G, Guellati S, Hilico L, Lambrecht A,
  Gu{\'e}rout R, Karr J~P, Nez F, Reynaud S, Szabo I~C, Tran V~Q, Trapateau J,
  Mohri A, Yamazaki Y, Charlton M, Eriksson S, Madsen N, Werf D, Kuroda N,
  Torii H, Nagashima Y, {Schmidt-Kaler} F, Walz J, Wolf S, Hervieux P~A,
  Manfredi G, Voronin A, Froelich P, Wronka S and Staszczak M 2014 {\em
  Hyperfine Interactions\/} {\bf 228} 141--150
  \urlprefix\url{https://doi.org/10.1007/s10751-014-1019-6}

\bibitem{Perez2015}
P{\'e}rez P, Banerjee D, Biraben F, {Brook-Roberge} D, Charlton M, Clad{\'e} P,
  Comini P, Crivelli P, Dalkarov O, Debu P, Douillet A, Dufour G, Dupr{\'e} P,
  Eriksson S, Froelich P, Grandemange P, Guellati S, Gu{\'e}rout R, Heinrich
  J~M, Hervieux P~A, Hilico L, Husson A, Indelicato P, Jonsell S, Karr J~P,
  Khabarova K, Kolachevsky N, Kuroda N, Lambrecht A, Leite A~M~M, Liszkay L,
  Lunney D, Madsen N, Manfredi G, Mansouli{\'e} B, Matsuda Y, Mohri A,
  Mortensen T, Nagashima Y, Nesvizhevsky V, Nez F, Regenfus C, Rey J~M, Reymond
  J~M, Reynaud S, Rubbia A, Sacquin Y, {Schmidt-Kaler} F, Sillitoe N, Staszczak
  M, {Szabo-Foster} C~I, Torii H, Vallage B, Valdes M, {Van der Werf} D~P,
  Voronin A, Walz J, Wolf S, Wronka S and Yamazaki Y 2015 {\em Hyperfine
  Interactions\/} {\bf 233} 21--27
  \urlprefix\url{https://doi.org/10.1007/s10751-015-1154-8}

\bibitem{Walz2004}
{Walz} J and {H{\"a}nsch} T~W 2004 {\em General Relativity and Gravitation\/}
  {\bf 36} 561--570
  \urlprefix\url{https://doi.org/10.1023/B:GERG.0000010730.93408.87}

\bibitem{Dufour2014}
Dufour G, Debu P, Lambrecht A, Nesvizhevsky V~V, Reynaud S and Voronin A~Y 2014
  {\em The European Physical Journal C\/} {\bf 74} 2731--
  \urlprefix\url{https://doi.org/10.1140/epjc/s10052-014-2731-8}

\bibitem{Hilico2014}
Hilico L, Karr J~P, Douillet A, Indelicato P, Wolf S and Schmidt-Kaler F 2014
  {\em International Journal of Modern Physics: Conference Series\/} {\bf 30}
  1460269 \urlprefix\url{https://doi.org/10.1142/S2010194514602695}

\bibitem{Sillitoe2017}
Sillitoe N, Karr J~P, Heinrich J, Louvradoux T, Douillet A and Hilico L 2017
  {\em $\bar{\text{H}}^{+}$ Sympathetic Cooling Simulations with a Variable
  Time Step\/} ({\em JPS Conf. Proc.\/} vol~18)
  \urlprefix\url{https://journals.jps.jp/doi/abs/10.7566/JPSCP.18.011014}

\bibitem{Radics2019}
Radics B, Janka G, Cooke D~A, Procureur S and Crivelli P 2019 {\em Review of
  Scientific Instruments\/} {\bf 90} 093305
  \urlprefix\url{https://doi.org/10.1063/1.5109315}

\bibitem{Wigner1932}
Wigner E 1932 {\em Phys. Rev.\/} {\bf 40}(5) 749--759
  \urlprefix\url{https://doi.org/10.1103/PhysRev.40.749}

\bibitem{Lykke1991}
Lykke K~R, Murray K~K and Lineberger W~C 1991 {\em Phys. Rev. A\/} {\bf 43}(11)
  6104--6107 \urlprefix\url{https://link.aps.org/doi/10.1103/PhysRevA.43.6104}

\bibitem{Vandevraye2014}
Vandevraye M, Babilotte P, Drag C and Blondel C 2014 {\em Phys. Rev. A\/} {\bf
  90} 013411
  \urlprefix\url{https://link.aps.org/doi/10.1103/PhysRevA.90.013411}

\bibitem{Bresteau2017}
Bresteau D, Blondel C and Drag C 2017 {\em Review of Scientific Instruments\/}
  {\bf 88} 113103 \urlprefix\url{https://doi.org/10.1063/1.4995390}

\bibitem{Frechet}
Fr\'echet M 1943 {\em Review of the International Statistical Institute\/} {\bf
  11} 182--205 \urlprefix\url{https://www.jstor.org/stable/1401114}

\bibitem{Cramer}
Cram\'er H 1999 {\em Mathematical Methods of Statistics (new edition)\/}
  ({Princeton University Press}) ISBN 978-0691005478

\bibitem{Refregier}
R{\'e}fr{\'e}gier P 2004 {\em Noise Theory and Application to Physics: From
  Fluctuations to Information\/} Advanced Texts in Physics ({New York}:
  {Springer}) ISBN 978-0-387-20154-2

\bibitem{Dufour2013}
Dufour G, G\'erardin A, Gu\'erout R, Lambrecht A, Nesvizhevsky V~V, Reynaud S
  and Voronin A~Y 2013 {\em Phys. Rev. A\/} {\bf 87}(1) 012901
  \urlprefix\url{https://link.aps.org/doi/10.1103/PhysRevA.87.012901}

\bibitem{GBARpage}
{{GBAR Collaboration}}  \urlprefix\url{https://gbar.web.cern.ch/}

\end{thebibliography}

\end{document}